\documentclass[aps,prb,amsmath,twocolumn,amssymb,floatfix,showpacs,superscriptaddress,nofootinbib,longbibliography,10pt]{revtex4-2}
\usepackage{float}
\usepackage{braket}
\usepackage[dvipsnames]{xcolor}
\usepackage{subfigure}
\usepackage[colorlinks=true,linktoc=page,linkcolor=blue,citecolor=blue,urlcolor=blue]{hyperref}
\usepackage{orcidlink}
\usepackage{comment}

\mathchardef\mhyphen="2D 

\newcommand{\non}{\nonumber}  
\allowdisplaybreaks

\makeatletter
\def\@fnsymbol#1{%
  \ensuremath{
    \ifcase#1
    \or \dagger
    \or \ddagger
    \or \mathsection
    \or \mathparagraph
    \or \|
    \or **
    \or \dagger\dagger
    \or \ddagger\ddagger
    \else
      \@ctrerr
    \fi
  }%
}
\makeatother

\begin{document}

\title{Observation of end-to-end pumping in a quasiperiodic Fibonacci-type photonic chain}

\author{Arnob~Kumar~Ghosh\textsuperscript{*}\,\orcidlink{0000-0003-0990-8341}}
\email{arnob.ghosh@physics.uu.se}
\affiliation{Department of Physics and Astronomy, Uppsala University, Box 524, 
751 20 Uppsala, Sweden}

\author{Ang~Chen\textsuperscript{*}}
\affiliation{Department of Physics, Stockholm University, S-10691 Stockholm, Sweden}

\author{Ashraf~El~Hassan}
\affiliation{Department of Physics, Stockholm University, S-10691 Stockholm, Sweden}

\author{{Patric~Holmvall}\,\orcidlink{0000-0002-1866-2788}}
\affiliation{Department of Physics and Astronomy, Uppsala University, Box 524, 
751 20 Uppsala, Sweden}

\author{Mohamed~Bourennane\,\orcidlink{0000-0002-6946-9996}}
\affiliation{Department of Physics, Stockholm University, S-10691 Stockholm, Sweden}

\author{{Annica~M.~Black-Schaffer}\,\orcidlink{0000-0002-4726-5247}}
\affiliation{Department of Physics and Astronomy, Uppsala University, Box 524, 
751 20 Uppsala, Sweden}

\begin{abstract}
End-to-end pumps utilizing edge-localized states offer a promising route to operate as connecting buses, supplying efficient and robust connectivity between non-neighboring elements in a network. Here, we investigate a finite quasiperiodic Fibonacci-type photonic chain and demonstrate its ability for end-to-end pumping, with only small and simple changes to the system. First, we use a tight-binding formalism to numerically show that a localized pumping state can be transferred between opposite ends of the system, with only a small structural change to the chain. Then, we experimentally implement this end-to-end pump in an array of coupled optical waveguides, where light propagation is effectively described by the tight-binding model under the paraxial approximation, enabling direct correspondence between theory and experiment. We numerically simulate and experimentally demonstrate pumping by injecting light into a single waveguide at one end of the setup, which activates a localized pumping state. As the light propagates along the wave guide array, it is also pumped to the other end. We further show that pumping remains robust against structural deformation, such as controlled defects in the waveguide array. Our results establish that quasiperiodic Fibonacci-type photonic lattices are a robust and experimentally viable platform for disorder-resilient state transfer.
\end{abstract}
\maketitle

\begingroup
\renewcommand\thefootnote{*}
\footnotetext{These authors contributed equally to this work.}
\endgroup

\setcounter{footnote}{0}

\section{Introduction}

The emergence of localized states at the boundaries of a system is a widespread phenomenon that arises in diverse physical contexts, including topological phases of matter, which host boundary states that are resilient to disorder and imperfections~\cite{hasan2010colloquium, qi2011topological, BernevigBOOK}.
Beyond electronic systems, topological phases have been extended to classical platforms, including in topological photonics, where synthetic lattices for light emulate topological phenomena~\cite{RechtsmanExperiment2013, LuNP2014, OzawaRMP2019}.
In particular, the analogy between Maxwell's equations for light under the paraxial approximation and the Schrödinger equation provides a playground for simulating condensed matter systems in photonic waveguide arrays~\cite{Christodoulides2003, LahiniPRL2008, LonghiQuantumOptics2009}. 
This gives direct experimental access to properties that may otherwise be challenging to probe in electronic materials.

\begin{figure}
\centering
\includegraphics[width=0.47\textwidth]{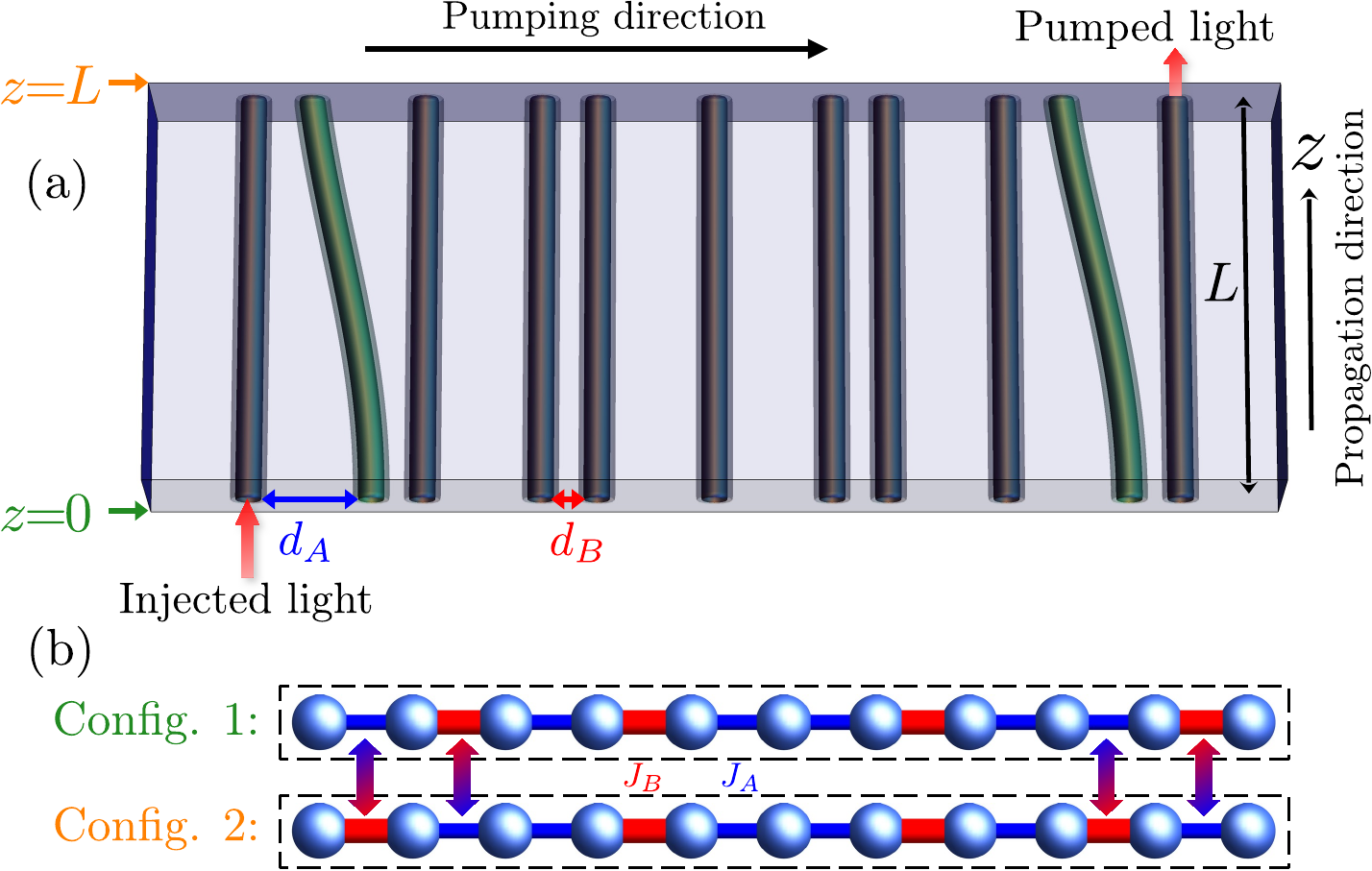}
\caption{(a) Schematic representation of the waveguide array illustrating the end-to-end pumping of light in a quasiperiodic Fibonacci-type system. The inter-waveguide separation has two characteristic distances $d_A$ and $d_B$ at $z=0$ and $L$. To pump the light, only two waveguides (green) are bent along the propagation direction ($z$). Light is then injected at $z=0$ at the left end (red arrow), and the pumped light emerges at $z=L$ from the opposite end of the array. (b) Tight-binding approximation of the waveguide array at $z=0$ (configuration 1) and $z=L$ (configuration 2), with weak bonds or coupling ($J_A$, corresponding to larger separation $d_A$) in blue and strong bonds ($J_B$, corresponding to smaller separation $d_B$) in red.
}
\label{Fig:Schematics}
\end{figure}

A particularly intriguing use of boundary states is adiabatic pumping, such as quantized charge pumping~\cite{ThoulessPRB1983, CitroNRP2023} and end-to-end pumping~\cite{KrausPumpingPRL2012, LonghiPRB2019}. While quantized charge pumping consists of a cyclic evolution of a Hamiltonian~\cite{ThoulessPRB1983} in a topological system, end-to-end pumping relies on the adiabatic switch between two Hamiltonians~\cite{LonghiPRB2019}, with both providing intriguing ways to transfer edge-localized states. In particular, end-to-end pumping has been explored in platforms such as the dimerized Su–Schrieffer–Heeger~(SSH) chain~\cite{LangNQI2017, MeiPRA2018, LonghiPRB2019, LonghiLandauZenerAQT2019, ZhengLiNaPRA2020, QiLuPRA2020, DAngelisPRR2020, PalaiodimopoulosPRA2021, CaoPRA2021, YuanAPLPh2021, QiPRRTR2021, WangPRA2022, ZhengLiNaYiPRApp2022, WeijiePRA2022, WangDaWeiPRA2023, ZhaonppjQuantumInf2023, Zurita2023fastquantumtransfer, RomeroPRApp2024, HanJinXuanPRApp2024, TianPRB2024, Fernandez2024, WangDaWeiPRA2024, ZhengCJP2025}, chiral edge states in quantum spin liquids~\cite{YaoNatComm2013, DlaskaQSI2017}, and even in quasiperiodic systems~\cite{KrausPumpingPRL2012, VerbinPump2015, SinghPRA2015, GhoshFC2025}. 
Eventually, end-to-end pumping has been experimentally demonstrated in a photonic waveguide array based on both the SSH chain model~\cite{WeijiePRA2022, ChaohuaPRA2023} and the Fibonacci quasicrystal~\cite{KrausPumpingPRL2012, VerbinPump2015}.

Despite these advances, a key practical limitation remains for end-to-end pumping.
The existing pumping schemes in SSH-based systems and quasicrystal implementations have all required continuous tuning of the bond strength along the entire length of the system~\cite{LangNQI2017, MeiPRA2018, LonghiPRB2019, LonghiLandauZenerAQT2019, ZhengLiNaPRA2020, QiLuPRA2020, DAngelisPRR2020, PalaiodimopoulosPRA2021, CaoPRA2021, YuanAPLPh2021, QiPRRTR2021, WangPRA2022, ZhengLiNaYiPRApp2022, WeijiePRA2022, WangDaWeiPRA2023, ZhaonppjQuantumInf2023, Zurita2023fastquantumtransfer, RomeroPRApp2024, HanJinXuanPRApp2024, TianPRB2024, Fernandez2024, WangDaWeiPRA2024, ZhengCJP2025, KrausPumpingPRL2012, VerbinPump2015, SinghPRA2015}. 
Such a need for exhaustive, fine-tuned control throughout the system increases the experimental complexity and inevitably introduces accumulated imperfections, especially as the system size increases. 
In contrast, a quasiperiodic setup has recently been theoretically shown to allow a significant reduction in the number of required tuning parameters~\cite{GhoshFC2025}, thereby simplifying end-to-end pumping as it requires only very few controls throughout the system. 
In this work, we exploit this advantage and demonstrate an efficient and simple end-to-end pumping scheme using a quasiperiodic photonic system, with only minor structural changes needed for pumping.

In particular, we consider a quasiperiodic system made from a Fibonacci-type photonic chain, formed by an array of coupled optical waveguides, as a platform for simple end-to-end pumping, see Fig.~\ref{Fig:Schematics}(a). 
The propagation of light in this waveguide setup can be effectively described by a tight-binding model under the paraxial approximation, see Fig.~\ref{Fig:Schematics}(b), enabling direct correspondence between theory and experiment.
Based on this framework, we first numerically simulate the end-to-end pumping of localized states and then experimentally demonstrate pumping by injecting light into a single waveguide at one end of the setup and detecting the light emerging from the opposite end via the intensity profile recorded by a CCD camera.
We also demonstrate that the pumping of light is robust against structural deformations, such as controlled defects in the waveguide array.

The remainder of this work is organized as follows. In Sec.~\ref{Sec:photonsetup}, we discuss the photonic setup, and in Sec.~\ref{Sec:TBmodel}, we discuss the tight-binding model for the photonic waveguide arrays. Section~\ref{Sec:End-to-endpump} is devoted to numerical demonstration of pumping based on tight-binding calculations. In Sec.~\ref{Sec:ExptResults}, we demonstrate the experimental pumping of light and compare it with the numerical simulations. In Sec.~\ref{Sec:Defects}, we show the stability of the end-to-end pumping with the structural defects. Finally, we conclude with a summary and outlook in Sec.~\ref{Sec:Summary}.

\section{Photonic setup} \label{Sec:photonsetup}
A quasiperiodic Fibonacci-type system contains quasiperiodic modulation of either hopping strength or on-site potential, following the Fibonacci sequence~\cite{KohmotoPRL1983, OstlundPRL1983, SireJPF1989, PiechonPRL1995, MadsenPRB2013, JagannathanRMP2021, RaiPRB2021}. This system has rapidly become a well-established platform to study the interplay between quasiperiodicity and a wide range of intriguing physical phenomena~\cite{KrausPRL2012, KrausNP2016, RaiPRB2019, MaceSciPost2019, RaiPRB2020, RaiPRB2021, ChiaracanePRB2021, MoustajNJP2023, SandbergPRB2024, KobialkaPRB2024, WangPRB2024, MarsalarXiv2025, BonselQuantum2026}, and experimentally realized in various settings~\cite{NegroPRL2003, SteurerJPDAP2007, KrausPumpingPRL2012, VerbinPRL2013, TanesePRL2014, VerbinPump2015, BabouxPRB2017, ZilberbergNature2018, LisieckiPRA2019, GoblotNatPhys2020, ReisnerPRB2023, FrancaPRBL2024}. In our photonic setup, we experimentally arrange waveguides in a linear array with nearest-neighbor separations belonging to the sequence $\left\{ d_j \right\}$ in Fig.~\ref{Fig:Schematics}(a).
The Fibonacci-type distribution $\left\{d_j\right \}$ is constructed using the sign of a Sturmian characteristic function 
\begin{align}
    \chi_j={\rm sgn}\left[\cos\left(2 \pi j \tau^{-1} + \pi \tau^{-1} + \phi\right) -\cos\left(\pi \tau^{-1}\right) \right],
\end{align}
with positive~(negative) values of $\chi_j$ setting the $j$-th distance to $d_B~(d_A)$~\cite{JagannathanRMP2021, VerbinPump2015}, representing short (long) separations between neighboring wave guides. Here, $\tau= (1+ \sqrt{5})/2$ is the golden ratio and $\phi \in [0, 2 \pi)$ the phason angle, which we use to tune the structure of the system~\cite{JagannathanRMP2021}.

The propagation of light in a classical waveguide array is described by Maxwell's equations, which under scalar and paraxial approximation becomes~\cite{Yariv1973, LonghiQuantumOptics2009}
\begin{align}
\small
i \partial_z \Psi(x,y,z) \!= \! \left[\! - \frac{1}{2k_0n_0} \! \nabla_\perp^2 \! - \! k_0 \Delta n(x,y) \right] \! \Psi(x,y,z).
\label{Eq:Paraxial}
\end{align}
Here $\nabla_\perp^2=\left( \partial_x^2 + \partial_y^2\right)$, $\Psi(x,y,z)$ is the amplitude of the electric field, $k_0=n_0 \omega /c$, with $\omega$ being the light frequency and $n_0$ the refractive index of the substrate glass, and $\Delta n(x,y)$ is the change in refractive index. 
In Appendix~\ref{App:Experimental_methods}, we detail the waveguide fabrication method in the experiment.
In the experiments, the separations were chosen as $d_A = 10~{\mu \rm m}$ and $d_B = 7~{\mu \rm m}$. The array of coupled optical waveguides for end-to-end pumping is schematically shown in Fig.~\ref{Fig:Schematics}(a).

\section{Tight-binding model} \label{Sec:TBmodel}
Equation~\eqref{Eq:Paraxial} for the light propagation in a classical waveguide array has a structure resembling the Schrödinger equation with the electric field $\Psi(x,y,z)$ and $z$ playing the role of wavefunction and time, respectively.
This analogy enables us to describe the waveguide setup using a `quantum' tight-binding Hamiltonian. This Hamiltonian consists of an onsite term, proportional to the eigenvalues of the principal mode of the waveguide, and a hopping term, proportional to the integral of the kinetic contribution~\cite{Christodoulides2003, LahiniPRL2008, LonghiQuantumOptics2009}. The onsite term depends on the waveguide's diameter, while the hopping term is determined by the separation between two waveguides~\cite{LahiniPRL2008}. In our case, we assume uniform waveguide diameters, yielding a constant onsite term that thus only produces a uniform shift in the system's spectrum. Thus, we can neglect the onsite term and keep only the hopping term as 
\begin{align}
    H=  \sum_{i,j \neq i}^N J_{ij} c_i^\dagger c_{j} + {\rm H.c.} ,
\label{Eq:HamHopping}
\end{align}
where $c_i^\dagger$ is the creation operator at lattice site $i$ and $J_{ij}$ is the bond strength, or coupling, between lattice site at $i$ and $j$, and $N$ is the total number of sites or equivalently waveguides. 
The bond strength depends on the distance between the waveguides, as $J_{ij} \simeq J e^{-d_{ij}/\xi}$, with $d_{ij}$ being the distance between two waveguides, $J$ the bare hopping amplitude, and $\xi$ the decay length of the coupling between two waveguides. 
In Appendix~\ref{SubsecApp:ExptCouplingCal}, we show experimentally measured changes in $J_{ij}$ as a function of distance between two waveguides and extract $J = 63.5~{\rm mm}^{-1}$ and $\xi = 2.2~{\mu \rm m}$. 
Since the hopping amplitude is exponentially suppressed between non-neighboring waveguides, we may effectively model the system with a nearest-neighbor tight-binding Hamiltonian using only two kinds of hoppings $J_A=Je^{-d_A/\xi}$ (weak) and $J_B=Je^{-d_B/\xi}$ (strong), as illustrated in Fig.~\ref{Fig:Schematics}(b). 
However, in all our numerical calculations, we keep $J_{ij}$ as the hopping between any two lattice sites, without resorting to the nearest-neighbor approximation.

\begin{figure}
\centering
\includegraphics[width=0.47\textwidth]{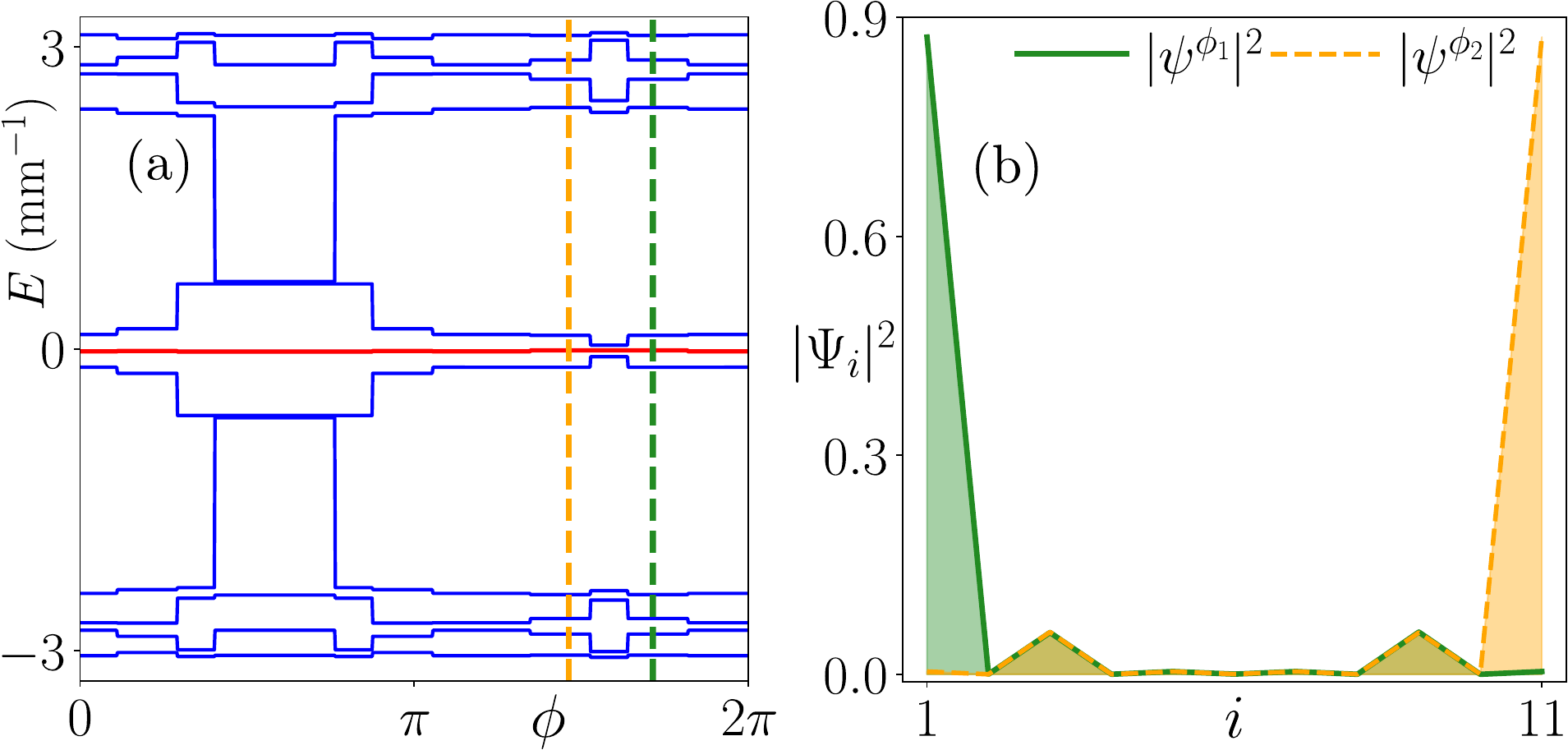}
\caption{(a) Eigenvalue spectra $E~({\rm mm}^{-1})$ of the Hamiltonian $H$ [Eq.~\eqref{Eq:HamHopping}] as a function of phason angle $\phi$. Red curve marks the pumping state. (b) Eigenstate of the pumping state as a function of the waveguide index $i$ for phason angles $\phi_1$ (green) and $\phi_2$ (orange).
}
\label{Fig:TBSpectrum}
\end{figure}

We analyze the eigenvalue spectra $E$ of the Hamiltonian in Eq.~\eqref{Eq:HamHopping} as a function of the phason angle $\phi$ in Fig.~\ref{Fig:TBSpectrum}(a). 
Varying the phason angle $\phi$ modifies the distance between neighboring waveguides and, as a result, the hopping sequence in the Hamiltonian $H$, thereby reshaping the spectrum. 
The red curve highlights the localized eigenstate we choose for pumping; we refer to this state as the pumping state. 
To facilitate end-to-end pumping, we choose two specific phason angles $\phi_1$ (configuration 1) and $\phi_2$ (configuration 2), denoted by green and orange dashed lines in Fig.~\ref{Fig:TBSpectrum}(a), respectively. 
The waveguide array and the corresponding hopping structure for these two phason angles are schematically illustrated in Figs.~\ref{Fig:Schematics}(a,b). 
As seen, to switch between configurations $1$ and $2$, we just need to change four hoppings (two hoppings per end), corresponding to bending only two wave guides, which makes our pump much simpler compared to ones based on the SSH chain or previous quasicrystal proposals~\cite{LangNQI2017, MeiPRA2018, LonghiPRB2019, LonghiLandauZenerAQT2019, ZhengLiNaPRA2020, QiLuPRA2020, DAngelisPRR2020, PalaiodimopoulosPRA2021, CaoPRA2021, YuanAPLPh2021, QiPRRTR2021, WangPRA2022, ZhengLiNaYiPRApp2022, WeijiePRA2022, WangDaWeiPRA2023, ZhaonppjQuantumInf2023, Zurita2023fastquantumtransfer, RomeroPRApp2024, HanJinXuanPRApp2024, TianPRB2024, Fernandez2024, WangDaWeiPRA2024, ZhengCJP2025, KrausPumpingPRL2012, VerbinPump2015, SinghPRA2015}.
In Fig.~\ref{Fig:TBSpectrum}(b), we show the distribution of the pumping state for phason angles $\phi_1$ (green) and $\phi_2$ (orange). 
Figure~\ref{Fig:TBSpectrum}(b) shows that the pumping state is localized on opposite ends for $\phi_1$ and $\phi_2$, a prerequisite for pumping \cite{GhoshFC2025}. We also verify the localized behavior of the pumping state experimentally, see Appendix~\ref{App:Calibration}. Figure~\ref{Fig:TBSpectrum} utilizes the phason angle $\phi$, a parameter arising from the quasiperiodic nature of the system, to identify the localized states. This demonstrates the crucial role of the system's quasiperiodic properties, encoded via the phason angle in designing the pump.

\section{End-to-end pumping scheme} \label{Sec:End-to-endpump}
To pump a state from one end of the system, we introduce a scheme that continuously interpolates between configurations $1$ and $2$ using the transfer protocol
\begin{align}
d_{1,9}(z) =& d_B - (d_B-d_A) \cos \left( \frac{\pi z}{2L}\right), \non \\
d_{2,10}(z) =& d_A + (d_B-d_A) \cos \left( \frac{\pi z}{2L}\right), \label{Eq:TransferProtocol}
\end{align}
where $L$ is the total propagation distance along $z$ and we introduced the shorthand notation $d_1 \equiv d_{12}$, $d_2 \equiv d_{23}$, etc. The transfer protocol for pumping state is schematically shown in Fig.~\ref{Fig:Schematics}(a). The protocol in Eq.~\eqref{Eq:TransferProtocol} involves bending only the second and tenth waveguides~\cite{GhoshFC2025}, which substantially simplifies the transfer protocol compared to previous quasiperiodic setups, where, in contrast, all waveguides/modes have to be controlled~\cite{KrausPumpingPRL2012, VerbinPump2015}. 
This waveguide bending amounts to modulating two outer bond strengths at each end of the system in the tight-binding Hamiltonian Eq.~\eqref{Eq:HamHopping}, see Fig.~\ref{Fig:Schematics}(b).
To ensure a successful end-to-end pumping, it is important to avoid level crossings during the protocol to maintain adiabaticity~\cite{LonghiPRB2019}. Throughout the pumping process, the pumping states' eigenvalues should remain separated from the remaining states by a so-called excitation gap, arising in this case from finite-size effects that depend solely on the number of sites/waveguides $N$~\cite{GhoshFC2025}.
Adiabaticity also depends on the light propagation distance $L$; a chain with a smaller excitation gap requires a longer $L$ for adiabaticity~\cite{LonghiPRB2019}. 
In the experimental setup, the propagation distance $L$ is limited by the physical dimensions of the waveguide array. 
Considering this experimental limitation on $L$, we choose $N=11$ as a good system size for a successful end-to-end pumping. In the main text, we focus on the total propagation length $L$, inter-waveguide separation distances, and the system size considered in the experimental setup, while we additionally show in Appendix~\ref{App:PhaseDiagram}  how the pumping fidelity and adiabaticity depend on these parameters. 

\begin{figure}
\centering
\includegraphics[width=0.49\textwidth]{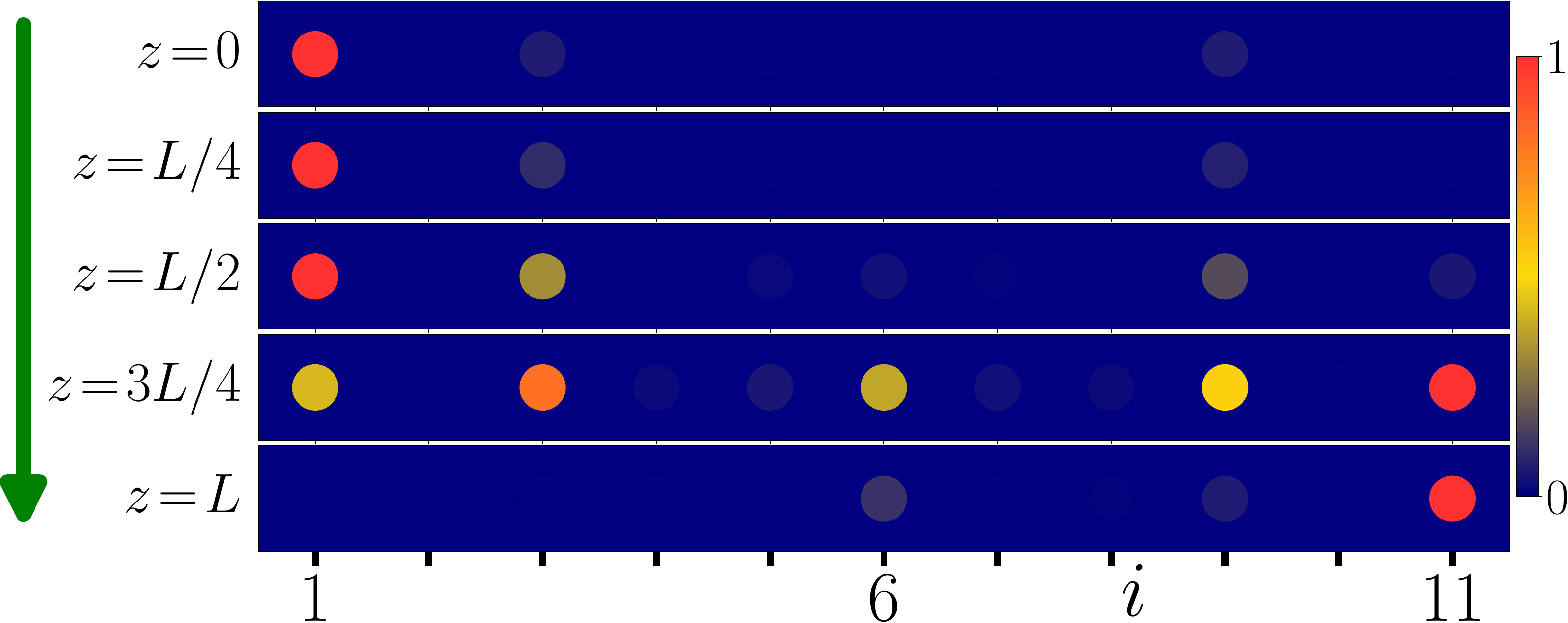}
\caption{Evolution of the pumping state at $z=0, L/4, L/2, 3L/4$, and $L$. Ticks on $x$-axes indicate the position of the waveguides and colorbar highlights the normalized weight of the propagating pumping state $\lvert \psi(z) \rvert^2$ at a given site. Here, $L=40~{\rm mm}$.
}
\label{Fig:TBpump}
\end{figure}

Next, we numerically demonstrate the end-to-end pumping with the transfer protocol Eq.~\eqref{Eq:TransferProtocol} and the Hamiltonian $H$ Eq.~\eqref{Eq:HamHopping}. 
We consider the pumping state in configuration $1$ (phason angle $\phi_1$, green state in Fig.~\ref{Fig:TBSpectrum}(b)) as the initial state $\ket{\psi(0)}$.
We obtain the instantaneous Hamiltonian $H(z)$ by replacing the hopping amplitudes $J_{1,2,9,10}(z)$ in Eq.~\eqref{Eq:HamHopping} with their $z$-dependent values in the transfer protocol Eq.~\eqref{Eq:TransferProtocol}. 
Here, we use shorthand notation $J_1 \equiv J_{12}$, $J_2 \equiv J_{23}$, etc. To obtain the propagating state $\ket{\psi(z)}$, we construct the propagation-evolution operator in a propagation-ordered~($\mathcal{T}$) manner as  $U(z,0)=\mathcal{T} \exp \left[ - i \int_0^z dz' H(z') \right] $, such that $\ket{\psi(z)}=U(z,0)\ket{\psi(0)} $.
Figure~\ref{Fig:TBpump} shows the numerical evolution of the pumping state at propagation distances $z=0, L/4, L/2, 3L/4,$ and $L$. At $z=0$, the pumping state is localized at the left end of the system. As this initial state evolves along the propagation direction, the state is adiabatically transported through the bulk, and at $z=L$ it becomes localized at the right end of the system. This numerically demonstrates that the pumping state can be transferred efficiently across the system. Thus, we obtain as the final state the pumping state for configuration $2$, i.e., the orange state in Fig.~\ref{Fig:TBSpectrum}(b).

We note that the pumping state is not a winding state of the Fibonacci quasicrystal~\cite{GhoshFC2025}, but rather a defect state, whose energy eigenvalue remains almost unchanged as we change the phason angle [see Fig.~\ref{Fig:TBSpectrum}(a)]. 
This defect state is similar to that observed in the SSH chain~\cite{WeijiePRA2022, ChaohuaPRA2023}, but it emerges instead in a quasiperiodic Fibonacci-type chain and, as such, requires only a much smaller tuning than in the SSH chain.
We choose this defect state over a winding state for pumping, as the defect state maintains a finite excitation gap throughout the pumping protocol, see Appendix~\ref{App:ExcitationGap}. The winding states have a smaller excitation gap and thus cannot easily maintain adiabaticity. Furthermore, in Appendix~\ref{App:SSHComparison}, we numerically compare our pumping scheme with an SSH-chain pump.

\section{Experimental demonstration of pumping} \label{Sec:ExptResults}
\begin{figure}
\centering
\includegraphics[width=0.49\textwidth]{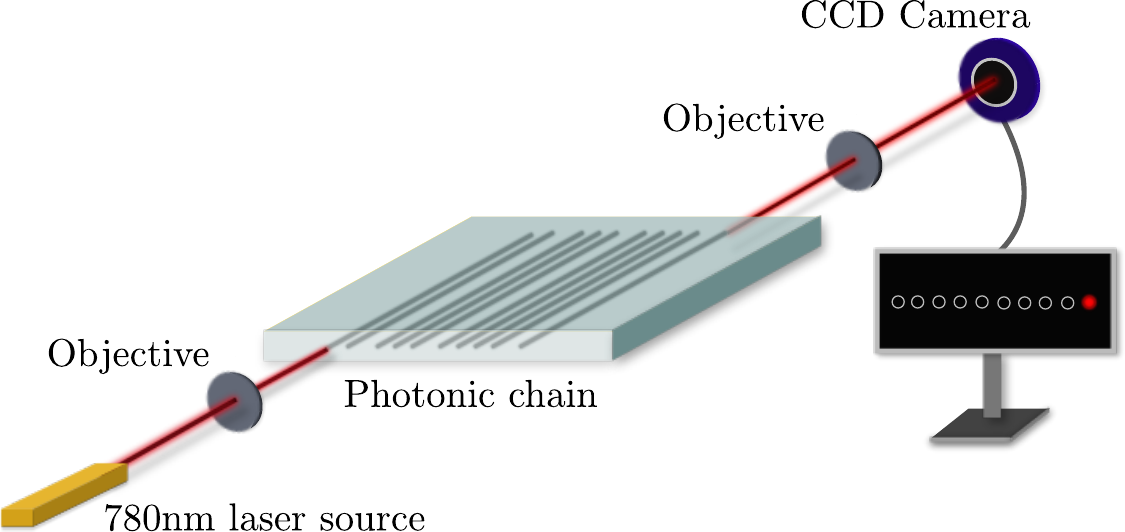}
\caption{Schematic of experimental setup. Light from a tunable laser is focused by an objective to selectively excite individual waveguides in the array. The output light is collected by a second objective and imaged onto a CCD camera.} 
\label{Fig:ExpSchematics}
\end{figure}

Having established end-to-end pumping numerically, we here establish experimental pumping of light in the waveguide array, see Fig.~\ref{Fig:ExpSchematics} for a schematic representation of the experimental setup.  
For optical characterization, light from a $780~{\rm nm}$ laser source is coupled to individual waveguides using a $10\times$ objective with a numerical aperture of $0.25$.
The initial excitation position is controlled to selectively address either edge or bulk waveguides. 
The sample output facet is imaged onto a CCD camera via another objective whose parameters match those of the input objective. 
The CCD camera enables direct measurement of the light-intensity distribution across the entire waveguide array. The recorded intensity profiles are used to analyze the result of the light propagation in the lattice. For each configuration studied in this work, two nominally identical samples were fabricated and measured to verify reproducibility.

The cosine bending profile of Eq.~\eqref{Eq:TransferProtocol} is experimentally implemented by an S-bend trajectory composed of two circular arc segments, which closely approximates the theoretical profile while remaining straightforward to fabricate (see Appendix~\ref{App:Experimental_methods} for details).

\begin{figure}
\centering
\includegraphics[width=0.4\textwidth]{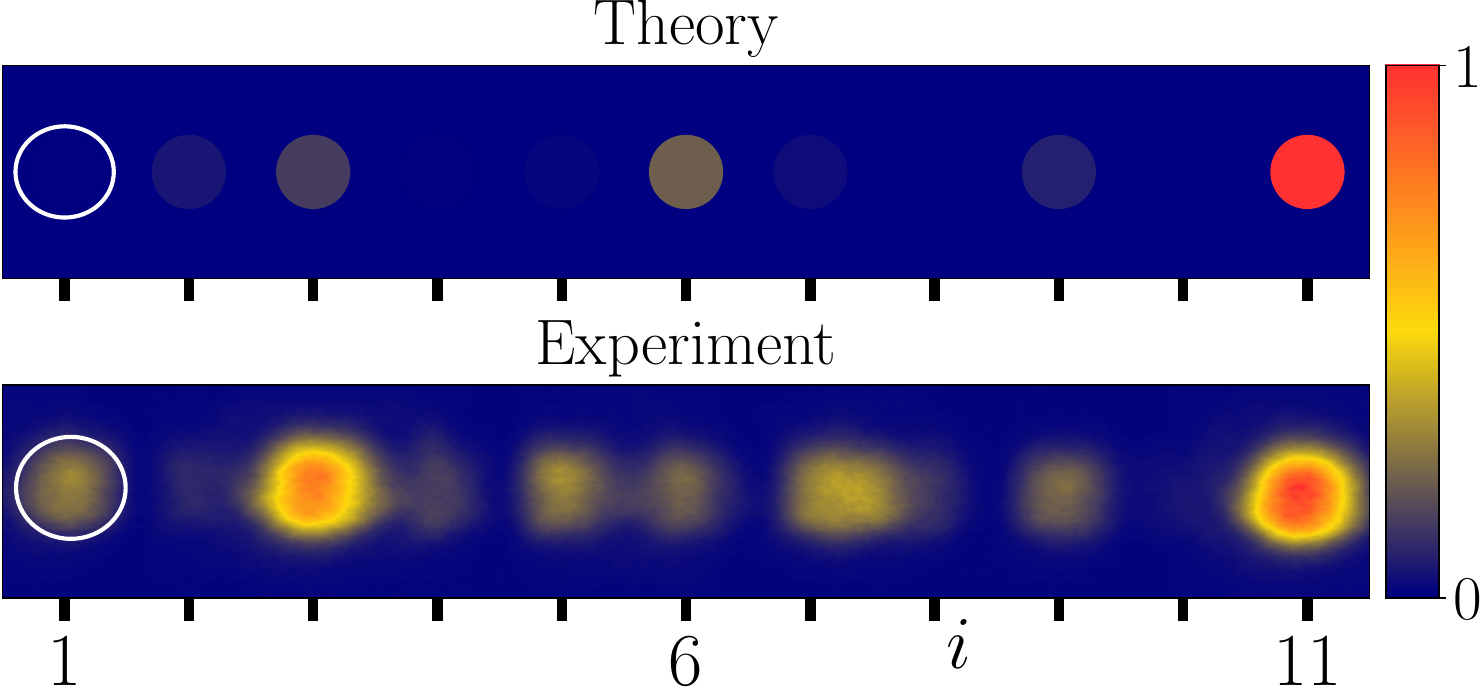}
\caption{Demonstration of end-to-end pumping of light in a waveguide array numerically (top panel, similar to bottom panel in Fig.~\ref{Fig:TBpump}) 
and experimentally (bottom panel). White circle indicates the injection point of light at $z=0$. Ticks on $x$-axes indicate the position of the waveguides and colorbar encodes the normalized light intensity at different waveguides at $z=L$.
}
\label{Fig:Pumping}
\end{figure}

In the waveguide setup, we do not have direct access to the eigenstate corresponding to the pumping state calculated in Fig.~\ref{Fig:TBpump}. 
Nevertheless, we can effectively imitate the pumping state, which is localized at the left end of the system, by injecting light only in the leftmost waveguide at $z =0$, see Appendix~\ref{App:Calibration}. 
Theoretically, we construct the initial state with a vector having only a non-zero element at the first site/waveguide of the system.
In Fig.~\ref{Fig:Pumping}, we show how the pumping state appears at $z=L$, after propagation in a wave guide following the transfer protocol Eq.~\eqref{Eq:TransferProtocol}, obtained both theoretically (top panel) and experimentally (bottom panel), where the white circle marks the position of the waveguide into which light is injected at $z=0$. Both theoretical and experimental results show that light emerges at the opposite end at $z=L$. This behavior provides proof-of-principle and confirms that the quasiperiodic Fibonacci-type waveguide setup adiabatically pumps a localized state between the system's opposite ends. 
In Appendix~\ref{App:LinearBending}, we also discuss the case when the inter-waveguide distances are changed linearly, instead of the transfer protocol Eq.~\eqref{Eq:TransferProtocol}, along the propagation direction.

\begin{figure}
\centering
\includegraphics[width=0.43\textwidth]{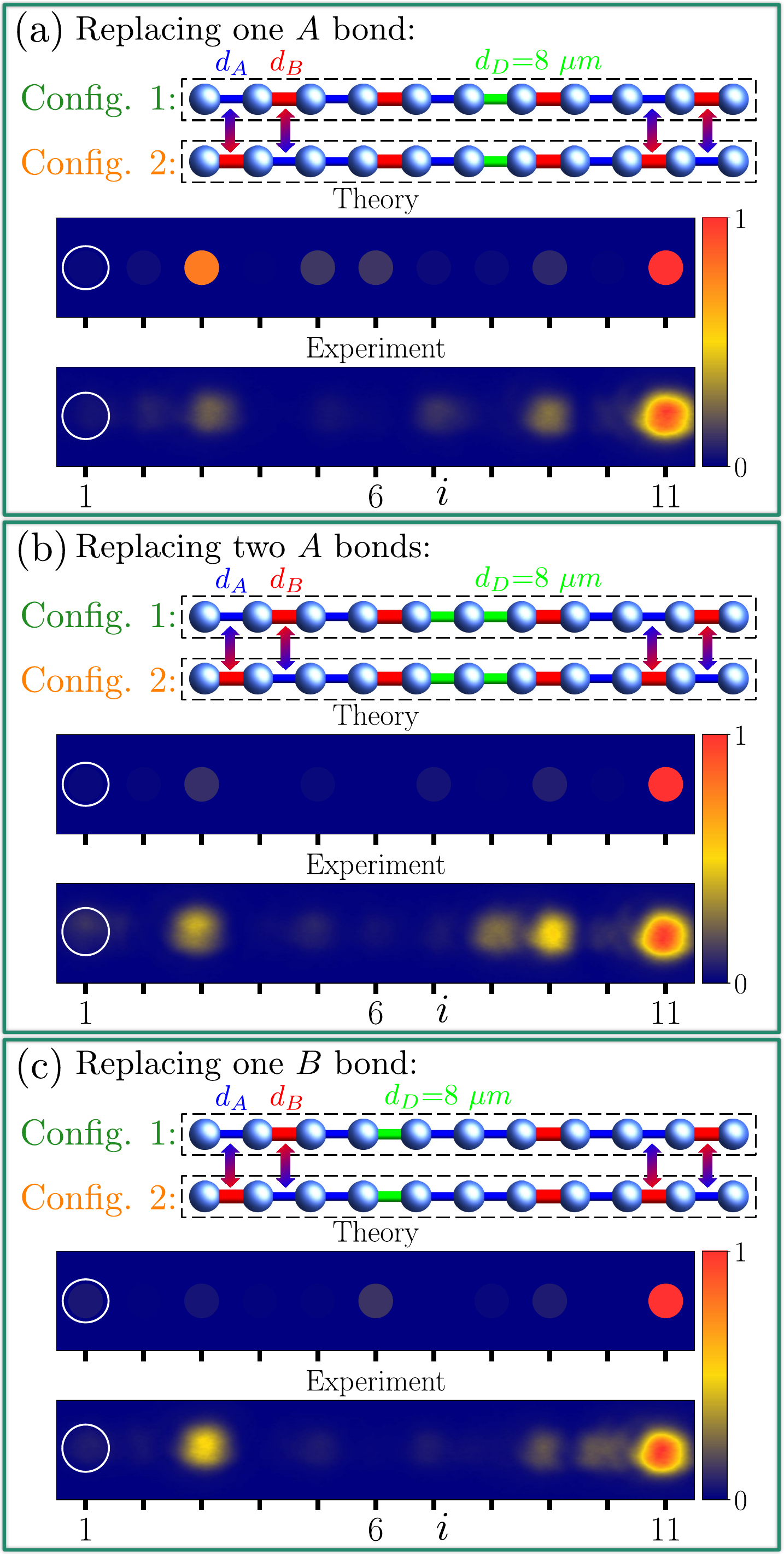}
\caption{Demonstration of pumping of light in a waveguide array in the presence of structural defects with (a) one $d_A$ distance replaced by $d_D$, (b) two $d_A$ distances are replaced by $d_D$, and (c) one $d_B$ distance is replaced by $d_D$. The top panels show a schematic picture of the waveguide array for configurations $1$ and $2$, with the defects represented by green, while the bottom two panels show the theoretical and experimental weights of the pumped light at $z=L$. White circle indicates the injection point of light at $z = 0$. Ticks on $x$-axes indicate the position of the waveguides and colorbar encodes the normalized light intensity in the different waveguides at $z = L$.
}
\label{Fig:Defect}
\end{figure}

\section{Stability against defects} \label{Sec:Defects}
We next investigate effect of structural defects in the waveguide array. 
A structural defect is introduced by locally modifying the distance between two neighboring waveguides such that it differs from $d_{A,B}=7,10~{\rm \mu m}$, here for concreteness setting the modified spacing by $d_D=8~{\rm \mu m}$. 
In Fig.~\ref{Fig:Defect}(a), we consider the case when one of the inter-waveguide distances $d_A$ is replaced by $d_D$, as illustrated in the schematic representation in the top panel with green bond indicating the defect. 
The bottom panels in Fig.~\ref{Fig:Defect}(a) show the corresponding theoretical and experimental results with this defect included. 
These results demonstrate that the light pumping persists in this system, despite the introduction of the structural defect. 
Moreover, the experimental data suggest that pumping is even more pronounced in the presence of the defect. 
We attribute this improved performance to the larger excitation gap induced by the structural defect, see Appendix~\ref{App:ExcitationGap} for details.

Next, we consider a configuration in which two of the inter-waveguide distances $d_A$ are replaced by $d_D$, see top panel in Fig.~\ref{Fig:Defect}(b). The bottom panels of Fig.~\ref{Fig:Defect}(b) present the corresponding theoretical and experimental results, demonstrating that the light pumping also persists in this case. Finally, we consider a case in which one single $d_B$ inter-waveguide distance is replaced with $d_D$, see top panel of Fig.~\ref{Fig:Defect}(c). The bottom panels again confirm the pumping of light. In Appendix~\ref{App:2Bbonds}, we showcase yet another case in which two of the inter-waveguide distances $d_B$ are replaced by $d_D$.
These results illustrate that pumping in the quasiperiodic Fibonacci-type chain does not rely on a fine-tuned setup but rather exhibits remarkable stability against structural defects. In fact, by studying the excitation gap in the presence of structural defects, we can even predict which defects can enhance the pumping characteristics, see Appendix~\ref{App:ExcitationGap}. Furthermore, in Appendix~\ref{App:Disorder}, we also numerically show that the end-to-end pumping remains robust even for rather substantial random structural disorder.

\section{Summary and outlook} \label{Sec:Summary}
In this work, we investigate a quasiperiodic Fibonacci-type chain as a platform for end-to-end pumping. 
First, we present theoretical results from a tight-binding model to demonstrate pumping in this quasiperiodic system. 
The pumping takes place via a defect state similar to that in the SSH chain, but our system requires considerably fewer modifications to achieve pumping.
Then, we realize the quasiperiodic Fibonacci-type chain in an optical waveguide array by tuning the spacing between the waveguides. 
We show experimentally that light injected into one end of the waveguide array is transported to the opposite end, demonstrating end-to-end light pumping via the localized state present in the system. 
Finally, we introduce controlled structural defects into the waveguide array, demonstrating the robustness of the pumping process and highlighting the system's resilience to imperfections. 
Our work thus provides a proof-of-principle of pumping in a one-dimensional quasiperiodic Fibonacci-type photonic lattice. 
Notably, this end-to-end pumping requires very minimal tuning in the setup, thereby dramatically simplifying the setup compared to the previous work based on SSH chains or quasicrystal setups~\cite{LangNQI2017, MeiPRA2018, LonghiPRB2019, LonghiLandauZenerAQT2019, ZhengLiNaPRA2020, QiLuPRA2020, DAngelisPRR2020, PalaiodimopoulosPRA2021, CaoPRA2021, YuanAPLPh2021, QiPRRTR2021, WangPRA2022, ZhengLiNaYiPRApp2022, WeijiePRA2022, WangDaWeiPRA2023, ZhaonppjQuantumInf2023, Zurita2023fastquantumtransfer, RomeroPRApp2024, HanJinXuanPRApp2024, TianPRB2024, Fernandez2024, WangDaWeiPRA2024, ZhengCJP2025, KrausPumpingPRL2012, VerbinPump2015, SinghPRA2015}.

Our work is also a stepping stone for many different intriguing future extensions. 
Here, we resort solely to a one-dimensional photonic lattice and illustrate the pumping of localized edge states. 
A natural extension is to construct a two-dimensional~(2D) photonic lattice by implementing the Sturmian sequence in both the $x$- and $y$-directions. 
The 2D system can support a higher-order topological phase hosting a corner mode~\cite{ElHassan2019} and thus provides a platform for pumping the corner state between different corners of the system.
Another promising future direction is to incorporate loss in the system, which can be effectively described by a non-Hermitian framework~\cite{SlootmanPRR2024}. 
Exploring the impact of loss on the pumping is of particular interest.
More intriguingly, it would be worthwhile to explore whether an engineered loss can improve the pumping process by enhancing the localization properties of the pumping state.

Moreover, in this work, we primarily consider a cosine-like or linear bending of waveguides along the propagation direction for pumping (see Fig.~\ref{Fig:Schematics}(a), Eq.~\eqref{Eq:TransferProtocol}, and Fig.~\ref{Fig:PumpingLinearApp}). 
A promising future direction is to improve the transfer protocol, enabling scaling to a larger system size by engineering a specific protocol within the framework of shortcuts to adiabaticity~\cite{GreentreePRB2004, GueryOdelinRMP2019, LiuPRL2025}.
Finally, our results indicate that the pumping performance can be enhanced in the presence of controlled structural defects. 
Consequently, future studies may exploit machine-learning-based approaches to identify an optimal waveguide-array configuration that further improves end-to-end pumping.

\subsection*{Data Availability}
The data that support the findings of this article are openly available at Ref.~\cite{ghosh_2026_21995226}.

\subsection*{Acknowledgments}
We thank Rodrigo Arouca for the insightful discussions. 
A.K.G., P.H., and A.M.B.-S. acknowledge financial support from the Swedish Research Council (Vetenskapsrådet) Grant No.~2022-03963, the Knut and Alice Wallenberg Foundation through the Wallenberg Academy Scholar program, KAW 2023.0244, and the European Union through the European Research Council (ERC) under the European Union’s Horizon 2020 research and innovation program (ERC-2022-CoG, Grant agreement No. 101087096). Views and opinions expressed are, however, those of the authors only and do not necessarily reflect those of the European Union or the European Research Council Executive Agency. Neither the European Union nor the granting authority can be held responsible for them.
A.C., A.E.H., and M.B. acknowledge financial support from the Swedish Research Council (Vetenskapsrådet) and the Knut and Alice Wallenberg Foundation. 
Part of the computations were enabled by resources provided by the National Academic Infrastructure for Supercomputing in Sweden (NAISS), partially funded by the Swedish Research Council through grant agreement No.~2022-06725.

\appendix

\section{Experimental methods} \label{App:Experimental_methods}
In this Appendix, we provide additional details on waveguide fabrication and coupling calibration to extract the needed parameters $J$ and $\xi$.

\subsection{Waveguide fabrication}

\begin{figure}[h]
	\centering
	\includegraphics[width=0.43\textwidth]{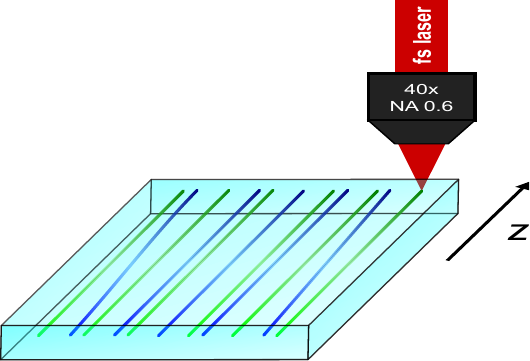}
	\caption{Schematic of the femtosecond laser direct-writing process. A $1030$ nm femtosecond laser is focused into the bulk glass to fabricate the waveguides by translating the sample along the $z$-direction.
	}
	\label{Fig:WaveguideFabrication}
\end{figure}

The waveguide arrays were fabricated using femtosecond laser direct writing in a transparent glass substrate. A femtosecond laser system (BlueCut fs laser from Menlo Systems) with a central wavelength of $1030~{\rm nm}$, a pulse duration of approximately $350$ fs, and a repetition rate of $1~{\rm MHz}$ was employed. The laser beam was focused into the bulk of the substrate using a $40\times$ objective with a numerical aperture~(NA) of $0.6$. 

During the fabrication process, the sample was mounted on a high-precision three-axis translation stage and translated at a constant speed of $40~{\rm mm/s}$, while the laser focus remained fixed. All waveguides were written at a depth of approximately $100~{\mu\rm m}$ below the surface to minimize surface-related aberrations and ensure homogeneous guiding conditions.

The writing laser power was optimized to produce single-mode waveguides at the probing wavelength. The fabricated waveguides support a single Gaussian mode at $780~{\rm nm}$, with a mode field diameter $(1/e^2)$ of approximately $10~{\mu\rm m}$. The resulting refractive index modification is estimated to be of the order of $\Delta n \sim 10^{-3}$. The propagation losses are estimated to be around $0.3~{\rm dB/cm}$.

To verify the reproducibility of our results, two nominally identical samples were fabricated for each configuration reported in this work, including the defect-free arrays (Figs.~\ref{Fig:Pumping}, ~\ref{Fig:Calibration}, and \ref{Fig:PumpingLinearApp}) and each of the structurally defected arrays (Figs.~\ref{Fig:Defect}, \ref{Fig:2Bdefect}, and \ref{Fig:PumpingDefectLinearApp}). Light injection and intensity measurements were performed independently on both samples under identical conditions.

\subsection{Coupling calibration} \label{SubsecApp:ExptCouplingCal}
The coupling constants between neighboring waveguides were experimentally calibrated using isolated two-waveguide directional couplers. For this purpose, pairs of identical waveguides with varying center-to-center separations were fabricated under the same writing conditions as the full lattice structures.

Light was then injected into one waveguide of each coupler and the power evolution along the propagation direction was characterized by measuring the output intensities at the end facet. The coupling strength $J_{ij}$ was then extracted from the measured power oscillations according to coupled-mode theory~\cite{Christodoulides2003, LahiniPRL2008, LonghiQuantumOptics2009}.

\begin{figure}
	\centering
	\includegraphics[width=0.45\textwidth]{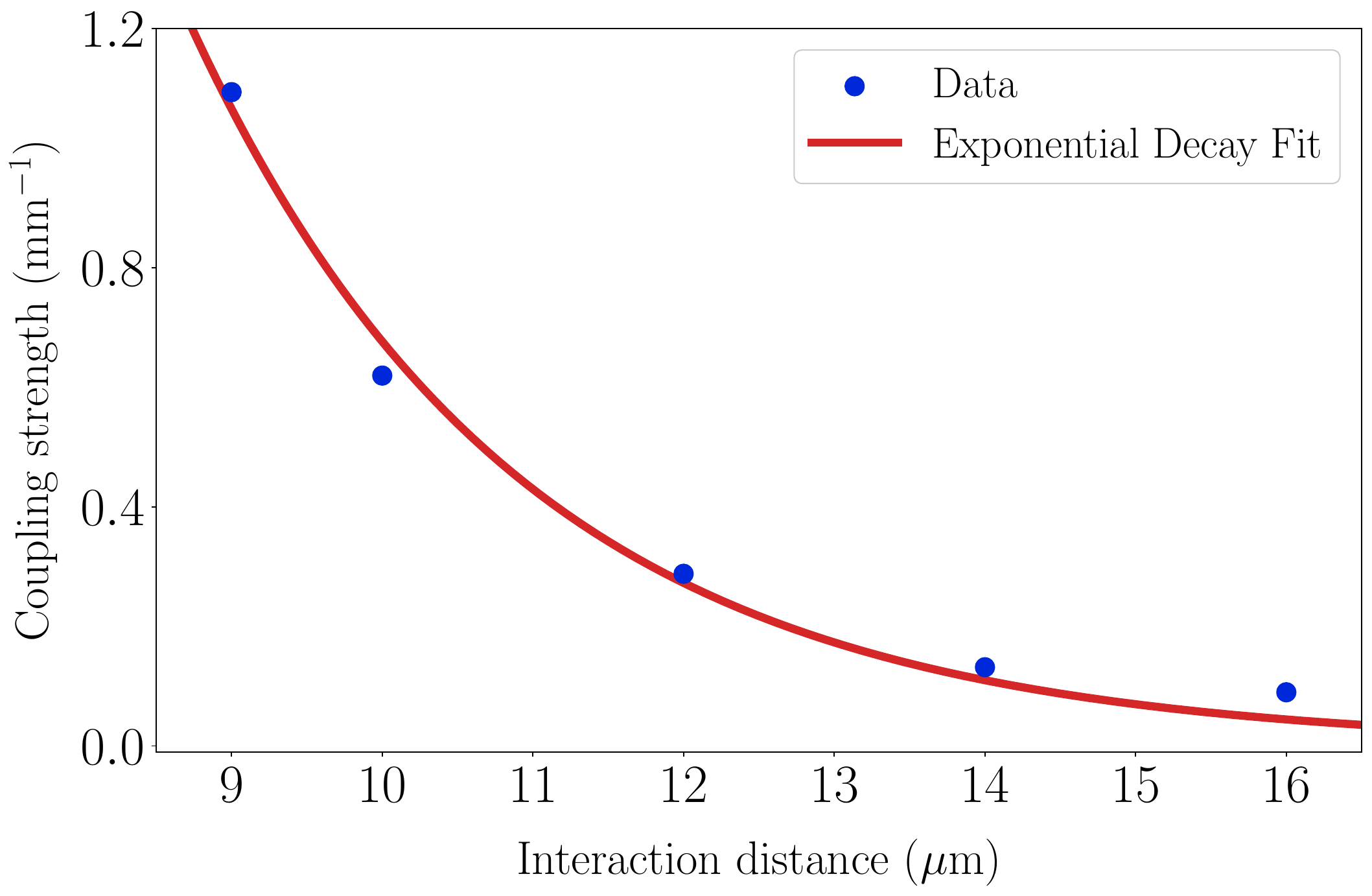}
	\caption{Measured coupling strengths (blue circles) and exponential decay fit (red line) as a function of the distance between two waveguides.  
	}
	\label{Fig:ExpHoppingCalibration}
\end{figure}

By repeating this procedure for different inter-waveguide separations, a quantitative relationship between the waveguide spacing and the coupling strength was obtained. The coupling strength between neighboring waveguides decreases approximately exponentially with their center-to-center separation $d_{ij}$: $J_{ij} \simeq  J e^{-d_{ij} / \xi}$. Using measured data, we extract coefficients $J = 63.5~{\rm mm}^{-1}$ , $\xi = 2.2~{\mu \rm m}$, with data and fit shown in Fig.~\ref{Fig:ExpHoppingCalibration}.

\subsection{Waveguide bending implementation}

In the experimental implementation, the bending of waveguides 2 and 10 is realized by an S-bend trajectory composed of two circular arc segments with opposite curvatures, joined at $z = L/2$. This trajectory 
shares the same endpoints, the same vanishing slope at $z = 0$ and $z = L$, and the same midpoint position as the cosine profile of Eq.~(3), but differs slightly in the intermediate region. In our experiment, the small bending angle, with arc radius $R = L^2/(4\Delta x) \approx 133$~m for $\Delta x = 3~\mu$m and $L = 40$~mm, ensures that bending losses are negligible. We emphasize that the precise functional form of the bending profile is not critical for end-to-end pumping: as long as the protocol smoothly interpolates between configurations 1 and 2 and remains adiabatic, pumping is preserved. This is further confirmed by the linear-bending results presented in Appendix.~\ref{App:LinearBending}.

\section{Localized pumping state} \label{App:Calibration}
\begin{figure*}
	\centering
	\includegraphics[width=0.75\textwidth]{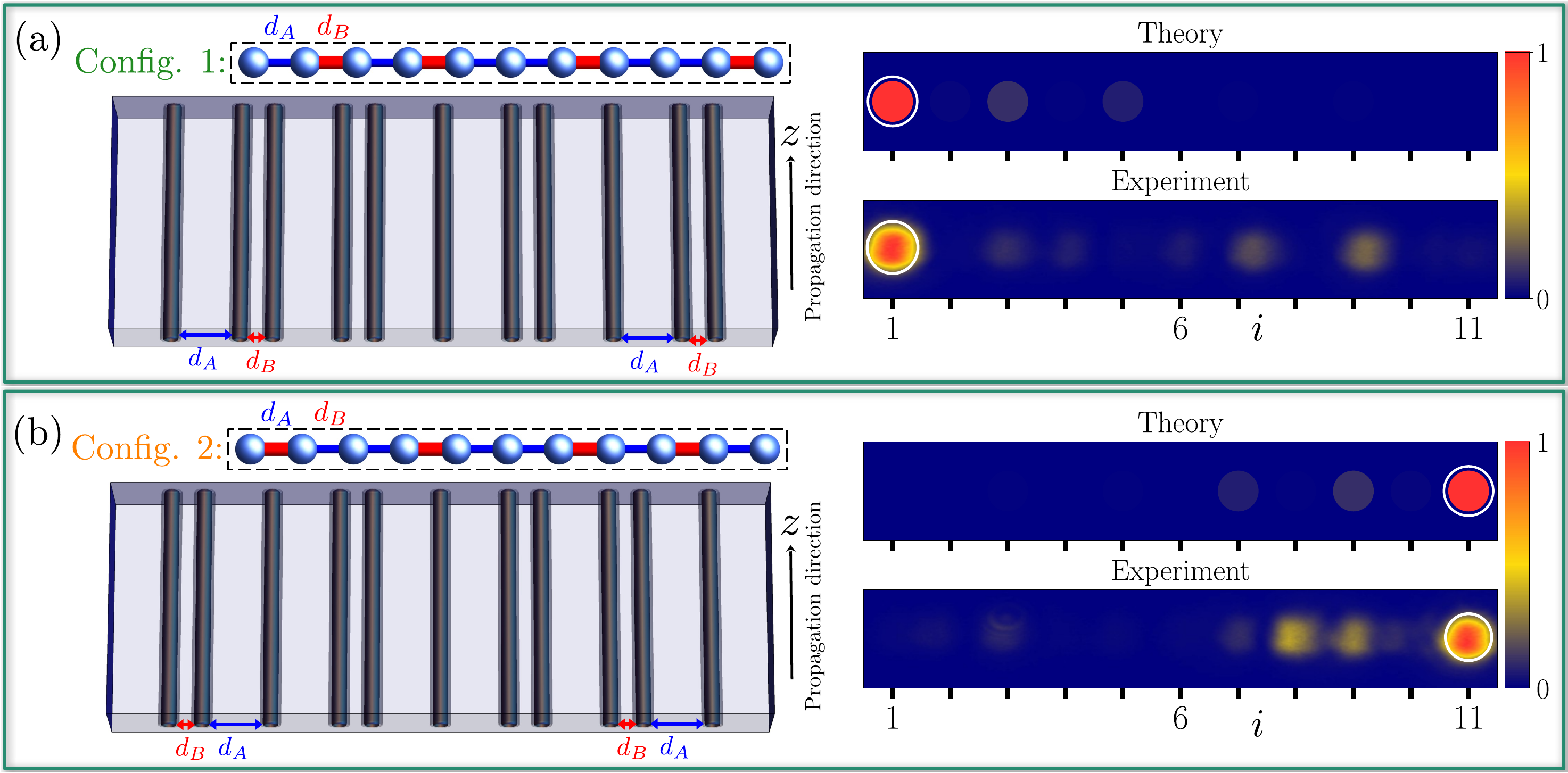}
	\caption{Demonstration of localized pumping state in the waveguide arrays for configuration $1$ (a) and $2$ (b), with configuration and waveguide arrays (left) and resulting light propagation (right). White circle indicates the injection point of light at $z = 0$. Ticks on $x$-axes indicate the position of the waveguides and colorbar encodes the normalized light intensity in the different waveguides at $z = L$.
	}
	\label{Fig:Calibration}
\end{figure*}

In this Appendix, we demonstrate that the localized pumping state can be effectively approximated by injecting light only into one of the outermost waveguides. To this end, we first consider the configuration $1$, see the left top panel of Fig.~\ref{Fig:Calibration}(a). The left bottom panel of Fig.~\ref{Fig:Calibration}(a) shows the corresponding waveguide array, which maintains constant distances along the propagation direction. For this configuration, the system supports a pumping state localized at the left end of the waveguide array. To emulate this pumping state, we inject light into the leftmost waveguide only, with the white circles in the right panel in Fig.~\ref{Fig:Calibration}(a) referring to this injection point at $z=0$. Owing to the presence of a localized pumping state, we observe that light remains confined to the same end of the waveguide array when it emerges on the other side of the waveguide at $z=L$, as demonstrated both theoretically and experimentally in Fig.~\ref{Fig:Calibration}(a), right panel. 

\begin{figure*}
	\centering
	\includegraphics[width=0.8\textwidth]{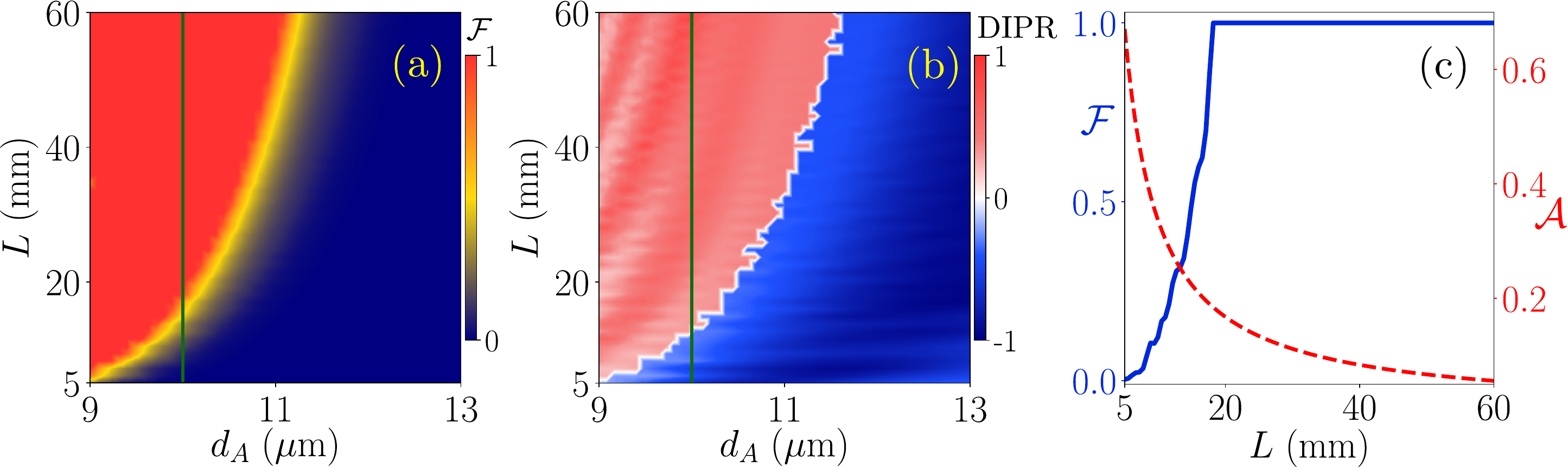}
	\caption{(a) Output fidelity $\mathcal{F}$ and (b) directional inverse participation ratio DIPR as a function of the inter-waveguide distance $d_A$ and total propagation distance $L$. We fix $d_B=7~\mu \rm m$. The green line indicates $d_A$ used in the main text and panel (c). (c) Output fidelity $\mathcal{F}$ as a function of total propagation distance $L$ with the right axis showing adiabaticity $\mathcal{A}$.
	}
	\label{Fig:PhaseDiagram}
\end{figure*}

Similarly, for configuration $2$, we have a right localized pumping state (left panels of Fig.~\ref{Fig:Calibration}(b)). To mimic this state, we inject light into the right-most waveguide at $z=0$ and demonstrate both theoretically and experimentally in Fig.~\ref{Fig:Calibration}(b), right panel, that the light remains localized at the same end at $z=L$.

\section{Phase diagram and parameter dependencies} \label{App:PhaseDiagram}

In this Appendix, we numerically explore how the end-to-end pumping depends upon the system parameters, especially the total propagation distance $L$ and the inter-waveguide distance $d_A$. We also discuss the effect of system size $N$ on the end-to-end pumping. For this purpose, we define the output fidelity $\mathcal{F}$ as
\begin{align}
    \mathcal{F}= \lvert\psi_{i=N}(z=L)\rvert^2/({\rm maximum}(\lvert\psi(z=L)\rvert^2)),
\end{align}
where $\psi(z=L)$ represents the final propagating state at $z=L$ and $\lvert\psi_{i=N}(z=L)\rvert^2$ denotes the weight of the final propagating state at the right-most waveguide with $i=N$. Thus, $\mathcal{F}=1$ indicates a state with full localization at the right end of the system, which is our target state, while $\mathcal{F}<1$ reveals that the final state does not have full localization at the right end.

We demonstrate the output fidelity $\mathcal{F}$ by varying the inter-waveguide distance $d_A$ and the total propagation distance $L$ while keeping $d_B=7~\mu \rm m$ in Fig.~\ref{Fig:PhaseDiagram}(a). From here, we find that for a lower inter-waveguide distance $d_A$, the output fidelity reaches unity (red region) for a shorter propagation distance. In comparison, for a larger $d_A$, a larger propagation distance is required to reach unit output fidelity. The green solid line in Fig.~\ref{Fig:PhaseDiagram}(a) indicates the $d_A$ chosen for all the numerical calculations and experimental setup in the main text.

We further supplement the fidelity phase diagram by computing the directional inverse participation ratio~(DIPR), defined as~\cite{ZengPRB2022}
\begin{align}
    {\rm DIPR}= {\rm sgn} \left[ \sum_{i}^{N} \left(i- \frac{N}{2}-\gamma \right) \lvert \psi_i(z=L) \rvert \right] {\rm IPR},
\end{align}
where the summation $i$ is over the number of sites and IPR stands for the inverse participation ratio, defined as ${\rm IPR}=\sum_i^N \lvert \psi_i(z=L) \rvert^4$. The constant $\gamma$ takes a value in the range $0<\gamma<0.5$ and we choose $\gamma=0.45$ in this work. The DIPR, unlike the conventional IPR, takes both positive and negative values, indicating right- and left-localized states, respectively. We demonstrate the DIPR in the $d_A \mhyphen L$ plane in Fig.~\ref{Fig:PhaseDiagram}(b), which indicates ${\rm DIPR}>0$, when output fidelity $\mathcal{F} \simeq 1$, while ${\rm DIPR}<0$ signifies that the final propagation state remains left localized, equivalently $\mathcal{F} \simeq 0$.

\begin{figure}
	\centering
	\includegraphics[width=0.49\textwidth]{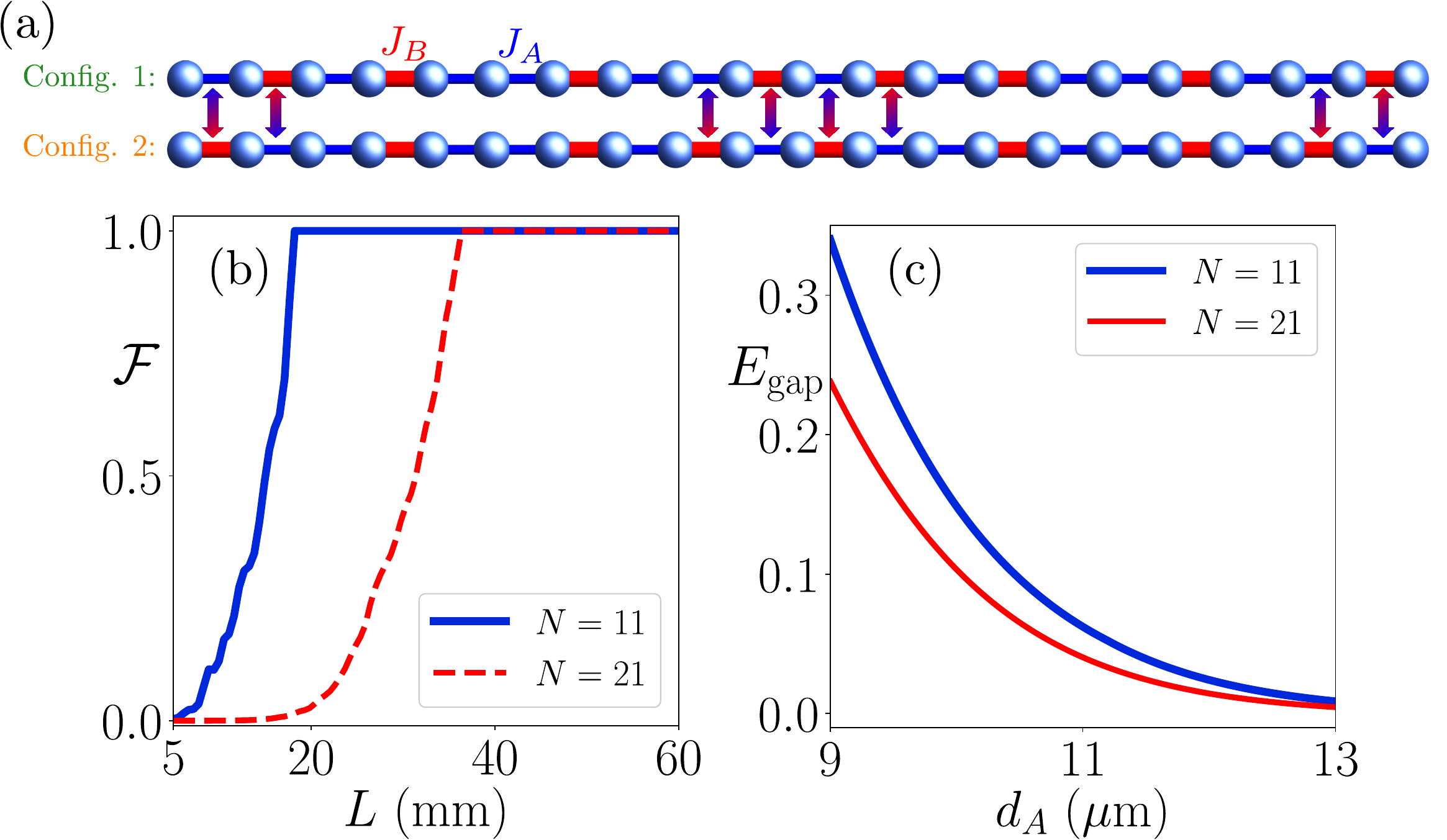}
	\caption{(a) Tight-binding approximation of the waveguide array for end-to-end pumping in $N=21$ system at $z=0$ (configuration 1) and $z=L$ (configuration 2), with weak bonds or coupling ($J_A$, corresponding to larger separation $d_A$) in blue and strong bonds ($J_B$, corresponding to smaller separation $d_B$) in red. (b) Output fidelity $\mathcal{F}$ as a function of the total propagation length $L$ for $N=11$ (blue) and $N=21$ (red) for a fixed inter-waveguide separation $d_A=10~\mu \rm m$ and $d_B=7~\mu \rm m$. (b) Minimum excitation gap $E_{\rm gap}$ as a function of the inter-waveguide separation $d_A$ for $N=11$~(blue) and $N=21$~(red) pump, for fixed $L=40~\rm mm$ and $d_B=7~\mu \rm m$. 
	}
	\label{Fig:FiniteSize}
\end{figure}

Next, we investigate the adiabatic condition for a fixed inter-wave guide distance $d_A$ by varying the total propagation distance $L$. To determine the adiabatic condition, we compute the following quantity~\cite{WeijiePRA2022}
\begin{align}
    \mathcal{A}(z)=\sum_{n \neq p} \frac{\braket{\psi_p(z)| \partial H(z) /\partial z| \psi_n(z)}}{E_n-E_p},
\end{align}
where $\psi_p(z)$ and $E_p$ represent the eigenstate and eigenvalue of the pumping state, and $H(z)$ is the instantaneous Hamiltonian as defined in the main text. For a successful pumping, $\mathcal{A}(z)$ should remain much smaller than one throughout the pumping period. Thus, instead of studying $\mathcal{A}(z)$, we study the maximum value of $\mathcal{A}(z)$ in a given pumping period, which we denote as $\mathcal{A} :=\max_{z \in [0,L]} \mathcal{A}(z)$. In Fig.~\ref{Fig:PhaseDiagram}(c), we plot $\mathcal{A}$ as a function of $L$ on the right axis and compare it with the output fidelity $\mathcal{F}$ on the left axis. As seen, $\mathcal{A}$ is large for a shorter total propagation distance $L$ and thus, the adiabatic condition is not satisfied, and, as a consequence, output fidelity is also close to zero. As we increase the total propagation distance $L$, the quantity $\mathcal{A}$ decreases, indicating that the pump approaches adiabaticity. However, since $\mathcal{A}$ is a slowly varying quantity, we also use the output fidelity $\mathcal{F}$ to determine whether pumping is successful or unsuccessful.

Finally, we compare the output fidelity $\mathcal{F}$ and excitation gap $E_{\rm gap}$ for different system sizes $N$. Here, the minimum excitation gap $E_{\rm gap}$ is defined as the smallest gap between the pumping state and the closest eigenstate when we continuously switch from one configuration to another. Note that $E_{\rm gap}$ is independent of the total propagation distance $L$, but only depends on the system size $N$ and the ratio between the inter-waveguide separation distances ($d_A$ and $d_B$)~\cite{GhoshFC2025}. To be able to utilize the same defect state that appears for $N=11$ in the main text, we repeat the system twice, such that the bigger system has a total of $N=2\times10+1=21$ sites or waveguides, see Fig.~\ref{Fig:FiniteSize}(a). Note that since we repeat the same unit cell twice, in the bigger system, we need to bend eight waveguides, or equivalently, we need to switch eight bonds to pump the localized state, as indicated by the bi-colored arrows in Fig.~\ref{Fig:FiniteSize}(a).

First, we show the output fidelity $\mathcal{F}$ as a function of the total propagation length $L$ in Fig.~\ref{Fig:FiniteSize}(b) for $N=11$ (blue) and $N=21$ (red). The plot shows that to reach $\mathcal{F}=1$, we need a longer total propagation length for a larger system size $N$. Thus, for our experimental setup, we choose the smaller $N=11$ to achieve faster state transfer from one end to the other. Furthermore, we also show how the minimum excitation gap $E_{\rm gap}$ depends on the inter-wave guide distance $d_A$ in Fig.~\ref{Fig:FiniteSize}(c) for $N=11$ (blue) and $N=21$ (red). The plot shows that the excitation gap is smaller for a bigger system size, and illustrates that choosing $N=11$ is the better choice: not only does it involve the least bent wave guides, but it also offers faster pumping. Finally, we note that it is also possible to use other odd-system sizes, but then one needs to redo calculations with those defect states. Overall, choosing $N=11$, given the experimental parameters and total propagation length limitation, is a good choice.

\section{Excitation Gap} \label{App:ExcitationGap}

\begin{figure*}
	\centering
	\includegraphics[width=0.8\textwidth]{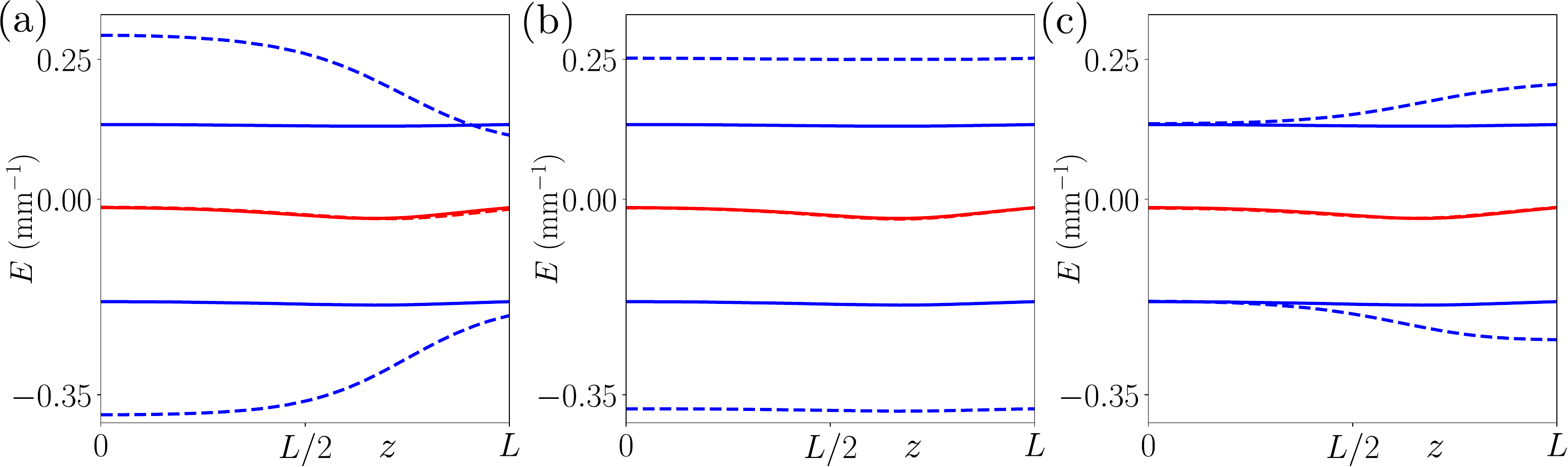}
	\caption{Eigenvalue of the pumping state (red) and its two closest eigenstates (blue) as a function of propagation distance $z$. Dashed curves represent the eigenvalues in the presence of a defect. Here, (a-c) correspond to the same defect configuration as in Fig.~\ref{Fig:Defect}(a-c) in the main text.
	}
	\label{Fig:ExcitationGapDefect}
\end{figure*}

In this Appendix, we discuss the excitation gap separating the localized pumping state from other states during the transfer process in the presence of bond defects. 
For this purpose, we focus on the eigenvalues of the pumping state and the two states closest to it, marked in red and blue, respectively, in Fig.~\ref{Fig:ExcitationGapDefect}. The dashed curve represents the eigenvalues in the presence of structural defects, with Figs.~\ref{Fig:ExcitationGapDefect}(a-c) corresponding to the same defect structure as in Figs.~\ref{Fig:Defect}(a-c) in the main text. 
The presence of a defect almost does not alter the eigenvalue of the pumping state (red). However, the other states (blue) actually move farther apart during the transfer process in the presence of a structural defect (dashed curves), especially in the presence of two bond defects [see Fig.~\ref{Fig:ExcitationGapDefect}(b)]. 
Owing to this enhancement of the excitation gap, the end-to-end pumping gets better in the presence of bond defects. Thus, by introducing appropriate structural defects into the system, we can engineer an improvement in pumping.

\section{Comparison with SSH-chain pump} \label{App:SSHComparison}

\begin{figure}
	\centering
	\includegraphics[width=0.49\textwidth]{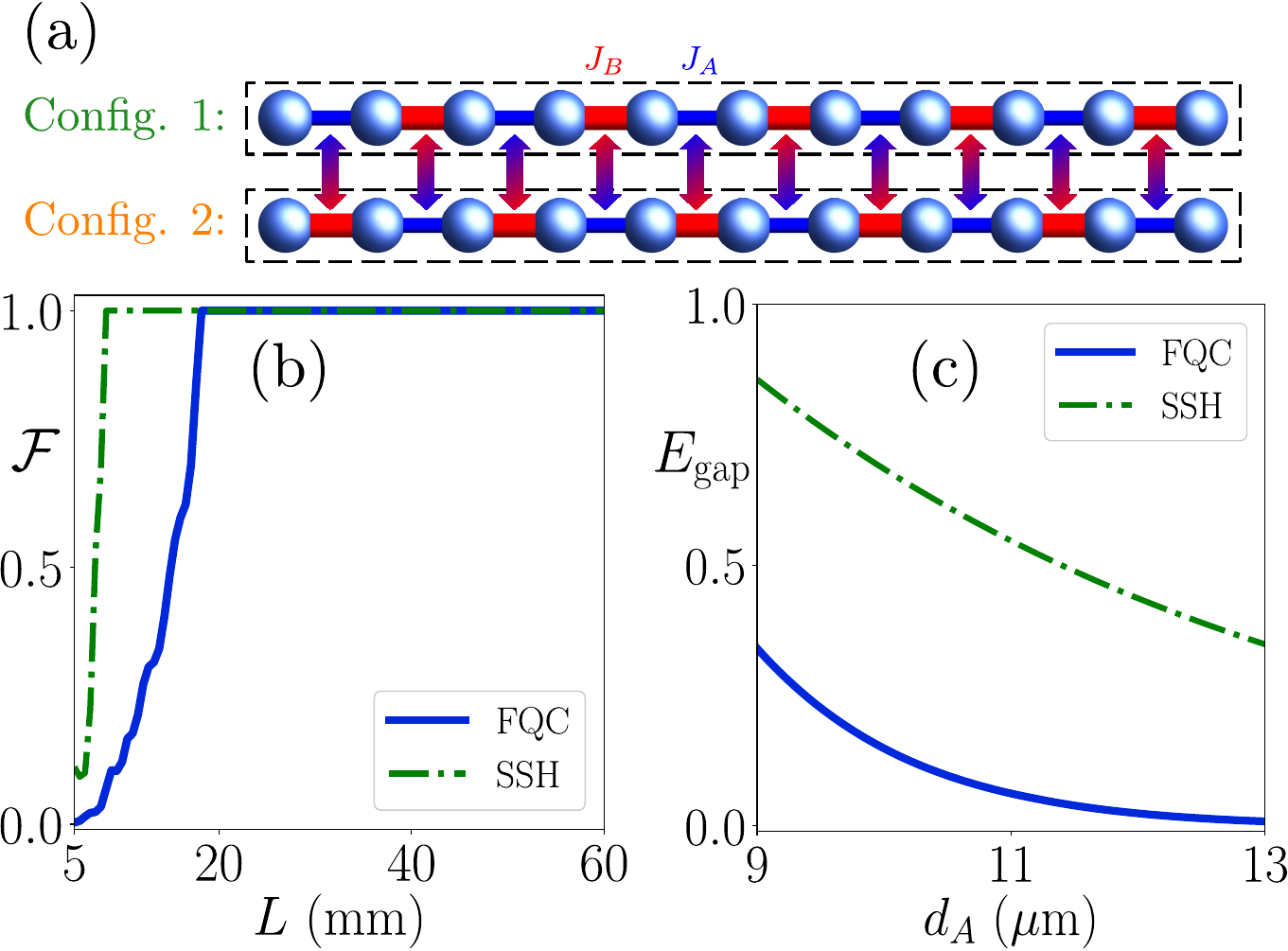}
	\caption{(a) Tight-binding approximation of the waveguide array for end-to-end pumping in the SSH-chain system at $z=0$ (Config.~1) and $z=L$ (Config.~2), with weak bonds or coupling ($J_A$, corresponding to larger separation $d_A$) in blue and strong bonds ($J_B$, corresponding to smaller separation $d_B$) in red. (b) Output fidelity $\mathcal{F}$ as a function of the total propagation length $L$ for SSH~(green) and FQC~(blue) pump, for a fixed value of $d_A=10~\mu \rm m$ and $d_B=7~\mu \rm m$. (c) Minimum excitation gap $E_{\rm gap}$ as a function of the inter-waveguide separation $d_A$ for SSH~(green) and FQC~(blue) pump, for a fixed $L=40~\rm mm$ and $d_B=7~\mu \rm m$.
	}
	\label{Fig:SSH}
\end{figure}

In this Appendix, we numerically discuss the end-to-end pumping in the SSH-chain. In an SSH chain, only two configurations are possible (for $J_A \neq J_B$), which we denote as configuration 1~(Config. 1) and configuration 2~(Config. 2). We illustrate these configurations in Fig.~\ref{Fig:SSH}(a). Config.~1~(2) supports a left~(right) localized mode. To pump a localized state in this SSH-based system, we need to continuously switch from Config.~1 to 2. Such switching between Config.~1 and 2, requires a flip of all bonds simultaneously (indicated by the bi-colored arrows in Fig.~\ref{Fig:SSH}(a)). Thus, as discussed in the main text, the SSH-based pump requires much more parameter tuning compared to the Fibonacci quasicrystal~(FQC) based pump.

Next, we compare the output fidelity $\mathcal{F}$ as a function of the total propagation distance $L$ for the SSH- and FQC-based pumps in Fig.~\ref{Fig:SSH}(b).  We observe that the output fidelity $\mathcal{F}$ reaches $1$ for a somewhat shorter total propagation distance $L$ with the SSH pump~(green) compared to the FQC pump~(blue). To understand the different minimum $L$ required for the SSH and FQC pumps, we investigate the minimum excitation gap $E_{\rm gap}$ as a function of the inter-waveguide distance $d_A$, while fixing $d_B=7~\mu \rm m$ in Fig.~\ref{Fig:SSH}(c). We see that the gap $E_{\rm gap}$ is larger for the SSH pump (green) when compared to the FQC pump (blue). As a result, in the SSH pump, we can pump the localized state already for a shorter total propagation distance $L$. However, the difference between the minimum propagation length for the two different systems is only around $10~{\rm mm}$ for the case studied in Fig.~\ref{Fig:SSH}(b). Thus, although the SSH-based pump can be slightly faster, our FQC-based pump still substantially simplifies the pumping process by requiring only a few parameters to be tuned, in contrast to the switching of all bonds required for the SSH pump.

\section{Linear bending} \label{App:LinearBending}

\begin{figure}
	\centering
	\includegraphics[width=0.49\textwidth]{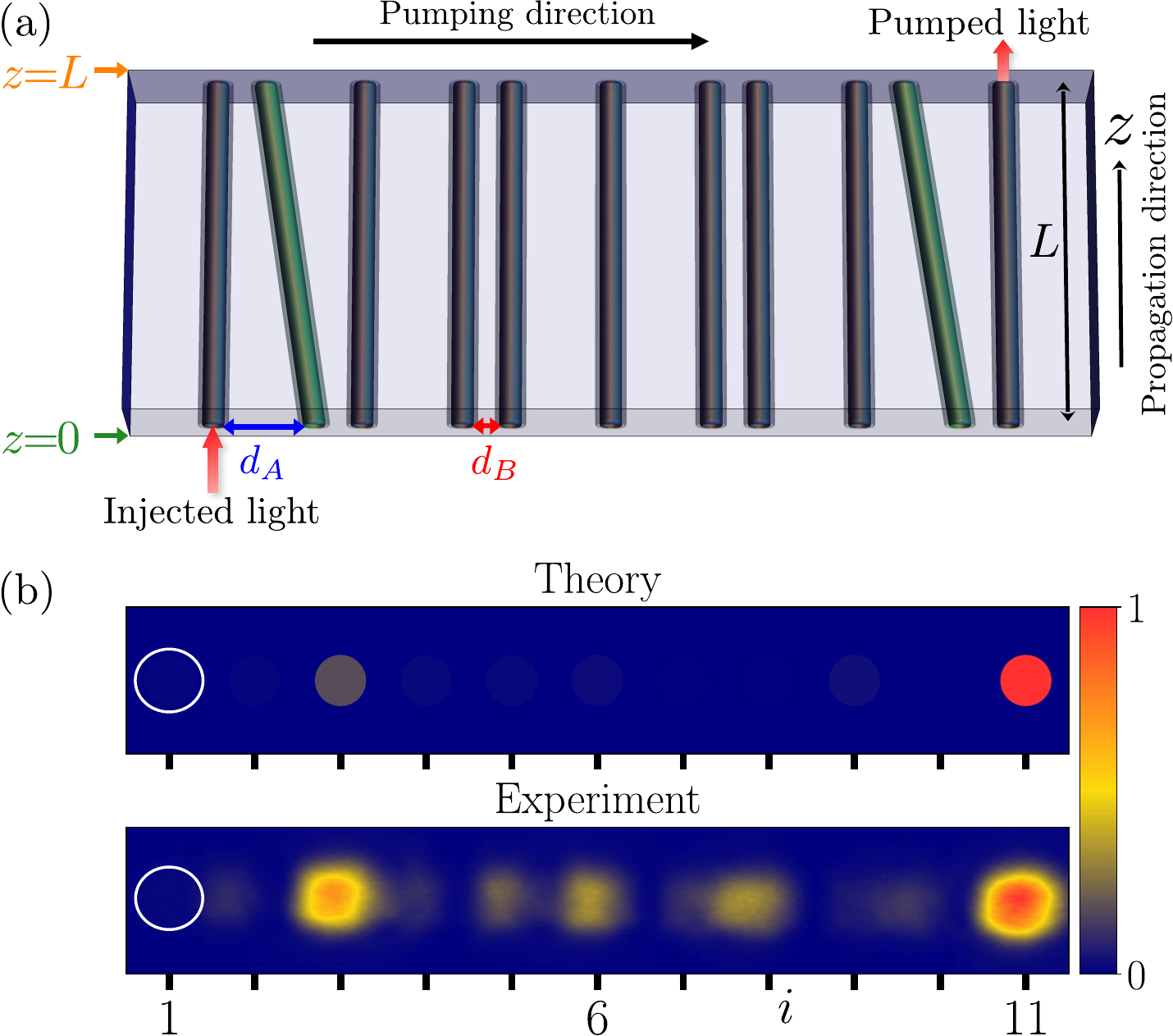}
	\caption{    
    (a) Schematic representation of the waveguide array for end-to-end pumping of light in the quasiperiodic Fibonacci-type system for the case of linear change of inter-waveguide distances. (b) Demonstration of end-to-end pumping of light in a waveguide array numerically (top panel) and experimentally (bottom panel). White circle indicates the injection point of light at $z=0$. Ticks on $x$-axes indicate the position of the waveguides and colorbar encodes the normalized light intensity in the different waveguides at $z=L$.
	}
	\label{Fig:PumpingLinearApp}
\end{figure}

In the main text, we analyze a particular bending profile of the waveguides following a cosine function, given by the transfer protocol in Eq.~\eqref{Eq:TransferProtocol}. In this Appendix, we present an alternative scenario in which the waveguides bend linearly and thus exhibit a linear change in the inter-waveguide distances. A schematic illustration of this configuration is shown in Fig.~\ref{Fig:PumpingLinearApp}(a), where the green waveguides are undergoing bending during the pumping process. Considering this linear bending, we show the end-to-end pumping of light both theoretically and experimentally in Fig.~\ref{Fig:PumpingLinearApp}(b), top and bottom panels, respectively, showing that light is pumped from the left end of the photonic chain to the right.  Overall, these results establish that light pumping remains effective also under linear bending and closely resembles the behavior reported in Fig.~\ref{Fig:Pumping} in the main text.

\begin{figure}
	\centering
	\includegraphics[width=0.4\textwidth]{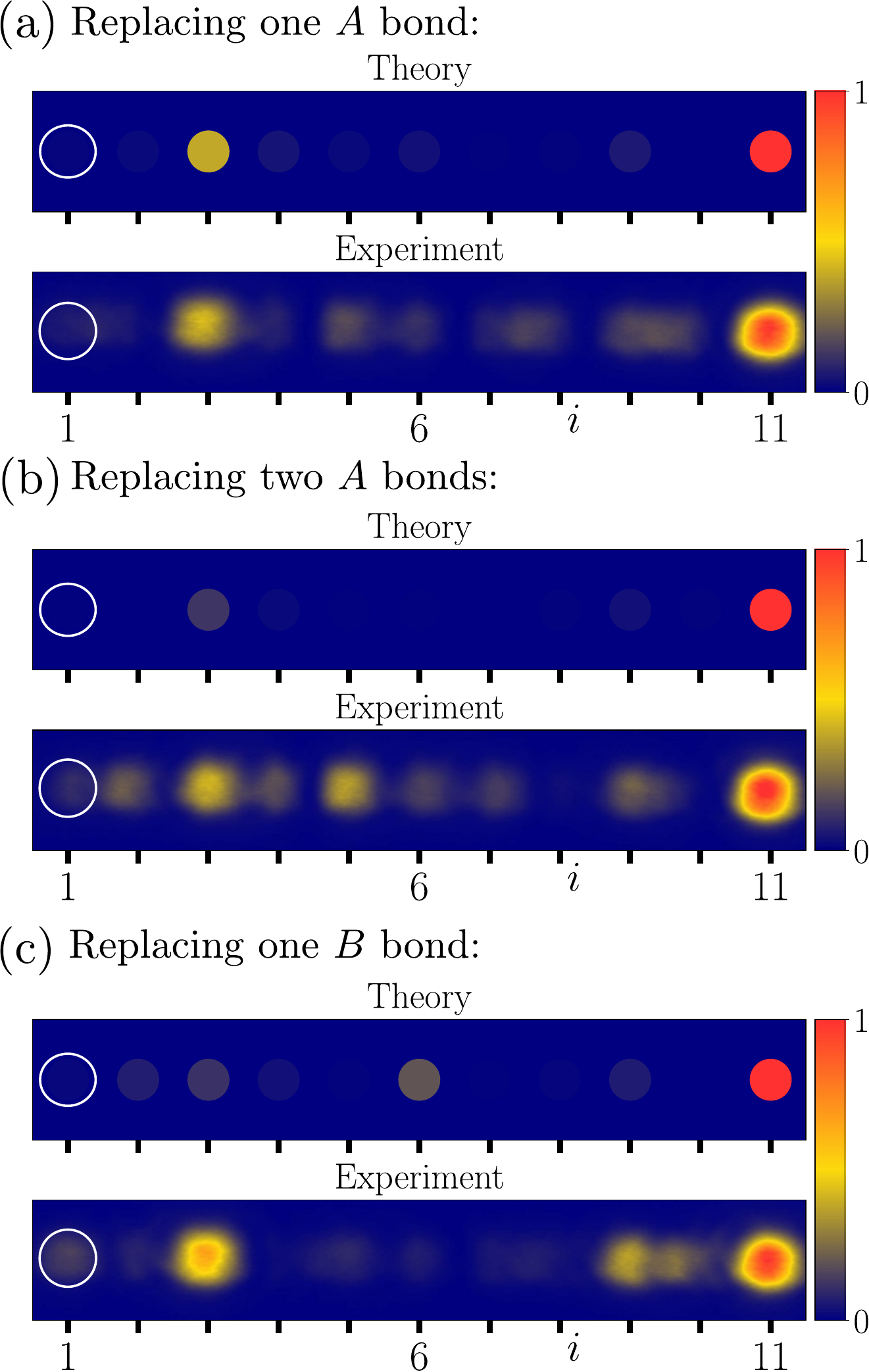}
	\caption{Demonstration of the pumping of light in a waveguide array in the presence of a structural defect with (a) one $d_A$ distance is replaced by $d_D$, (b) two $d_A$ distances are replaced by $d_D$, and (c) one $d_B$ distance is replaced by $d_D$. Thus, (a-c) correspond to the same configurations as Fig.~\ref{Fig:Defect}(a-c) in the main text, but for a linear bending of the wave guides. White circle indicates the injection point of light at $z=0$. Ticks on $x$-axes indicate the position of the waveguides and colorbar encodes the normalized light intensity in the different waveguides at $z=L$.
	}
	\label{Fig:PumpingDefectLinearApp}
\end{figure}

We also examine the effects of structural defects under linear bending, using the same defects as in Fig.~\ref{Fig:Defect} in the main text and show the result of end-to-end pumping in Fig.~\ref{Fig:PumpingDefectLinearApp}. Figure~\ref{Fig:PumpingDefectLinearApp}(a) shows the case where a single $d_A$ distance is replaced by $d_D$ and demonstrates the end-to-end pumping both theoretically and experimentally. Likewise, in Fig.~\ref{Fig:PumpingDefectLinearApp}(b) we depict the pumping when two $d_A$ distances are replaced by $d_D$, while Fig.~\ref{Fig:PumpingDefectLinearApp}(c) corresponds to the scenario when one $d_B$ distance is replaced by $d_D$. 
Thus, in this Appendix, we establish that end-to-end pumping of light works in the photonic waveguide setup, including in the presence of structural defects, also for linear bending of the waveguides.

\begin{figure}
	\centering
	\includegraphics[width=0.3\textwidth]{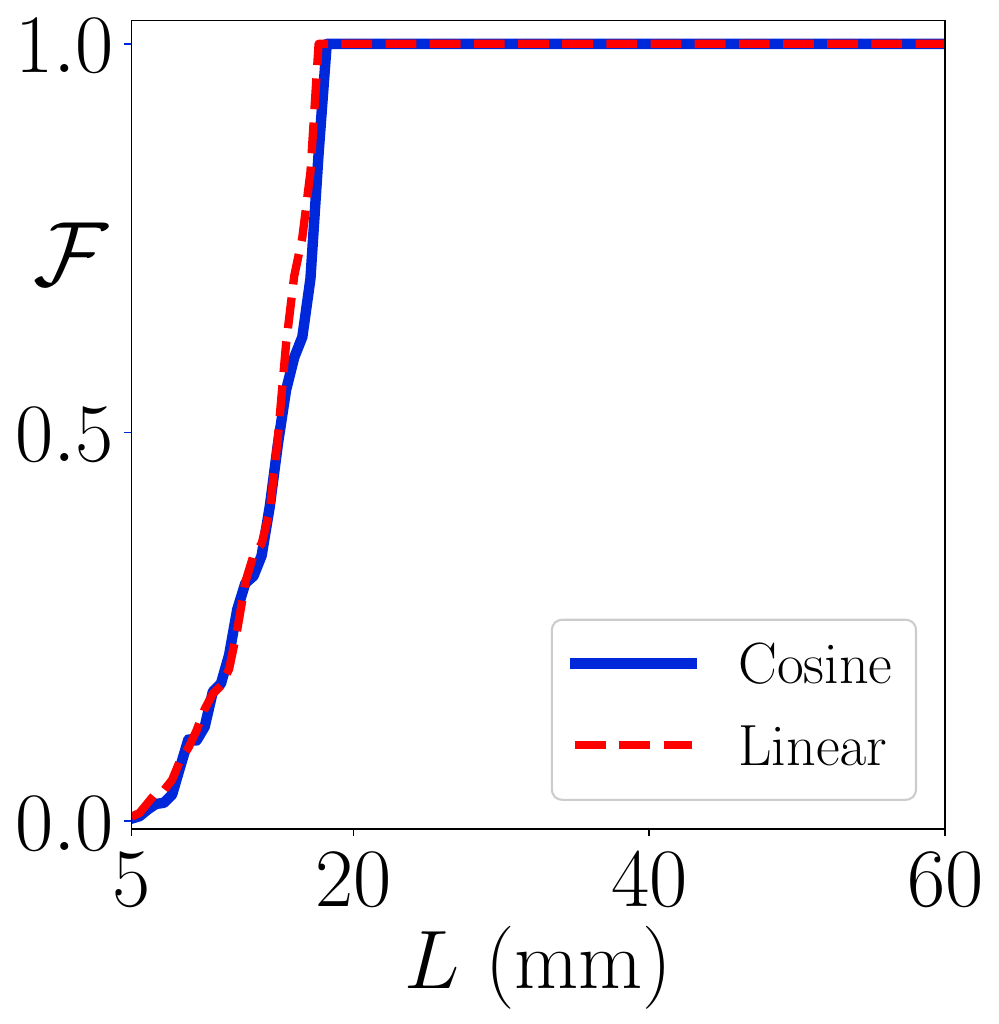}
	\caption{Comparison of the output fidelity $\mathcal{F}$ as a function of the total propagation length $L$ for the cosine- (blue) and linear-bending (red) scheme. Here $d_A=10~\mu \rm m$ and $d_B=7~\mu \rm m$.
	}
	\label{Fig:FidLinearComp}
\end{figure}

Finally, we compare the theoretical output fidelity $\mathcal{F}$ for the cosine- (blue) and linear-bending (red) scheme in Fig.~\ref{Fig:FidLinearComp} as a function of the total propagation length $L$. From Fig.~\ref{Fig:FidLinearComp}, we conclude that the output fidelity $\mathcal{F}$ behaves almost identically for both bending schemes.

\section{Replacing 2$B$ bonds} \label{App:2Bbonds}

\begin{figure}
	\centering
	\includegraphics[width=0.49\textwidth]{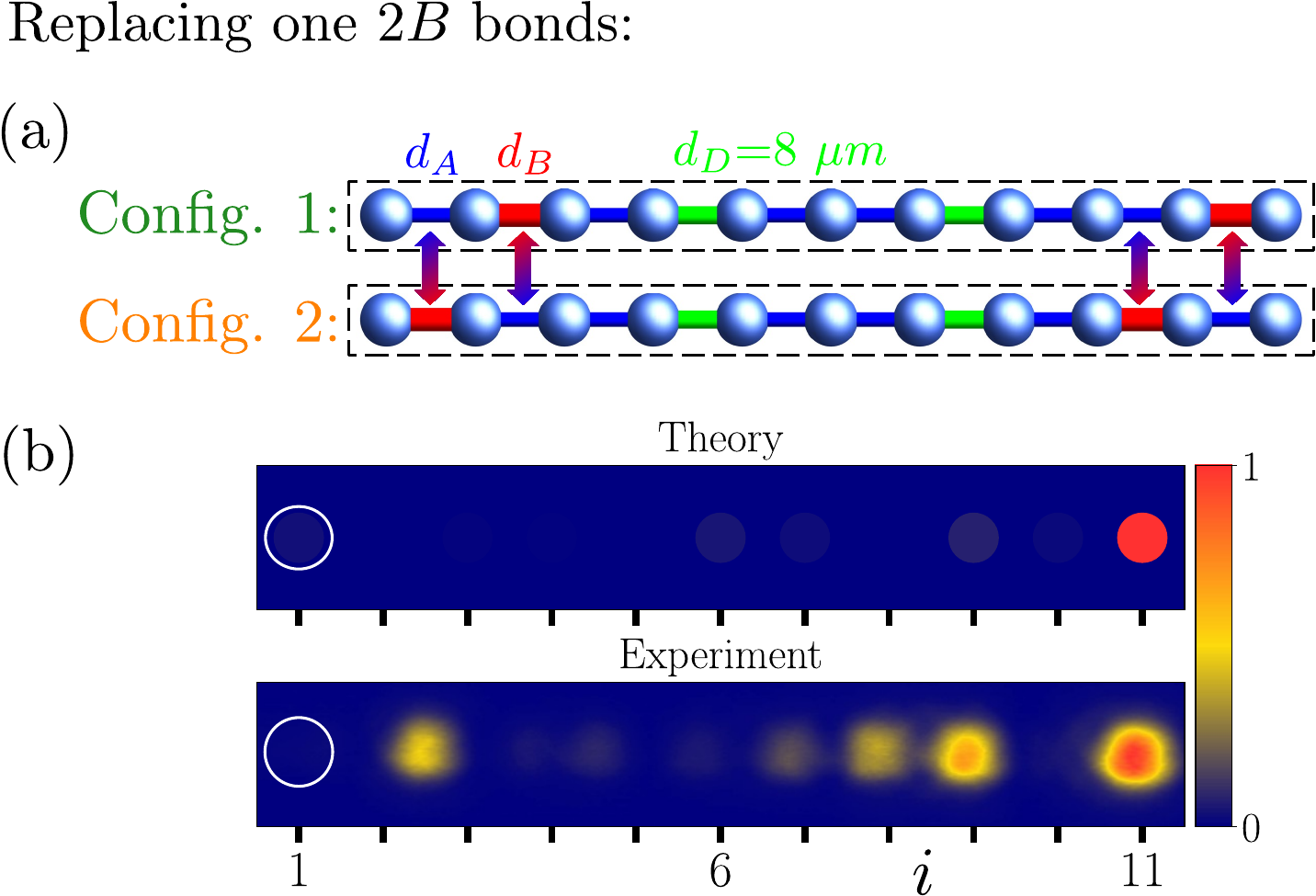}
	\caption{Demonstration of the pumping of light in a waveguide array in the presence of a structural defect consisting of two spatially separated $d_B$ distances replaced by $d_D$. (a) Schematic picture of the waveguide array for configurations $1$ and $2$, with the defects represented by green. (b) Theoretical and experimental weights of the pumped light at $z=L$. Ticks on $x$-axes indicate the position of the waveguides and colorbar encodes the normalized light intensity in the different waveguides at $z=L$.
	}
	\label{Fig:2Bdefect}
\end{figure}

In the main text, we investigate the effect of structural defects on end-to-end pumping. 
In this Appendix, we discuss another case in which we replace two separated inter-waveguide distances $d_B$ with $d_D$. 
Figure~\ref{Fig:2Bdefect}(a) shows a schematic representation of this setup with the green bonds representing the defects. 
In Fig.~\ref{Fig:2Bdefect}(b), we show the results for end-to-end pumping for both theory and experiment in the presence of these defects. The results illustrate that the end-to-end pumping remains robust in the presence of these defects. This section supplements the results presented in Fig.~\ref{Fig:Defect} of the main text for different structural defects.

\section{Random structural disorder} \label{App:Disorder}

In this Appendix, we discuss the effect of random structural disorder on the end-to-end pumping in our setup. For this, we introduce random fluctuations in the inter-waveguide separation through $d_{ij} \rightarrow \left( d_{ij}+ W_{ij} \right)$, where $W_{ij}$ represents the disorder, which we assume to be uniformly distributed as $W_{ij} \in [-W/2,W/2]$. Here, $W$ denotes the strength of the disorder. We compute the output fidelity $\mathcal{F}$ for different random disorder configurations and then take an average over the number of disorder configurations for each $W$.

\begin{figure}[h]
	\centering
	\includegraphics[width=0.49\textwidth]{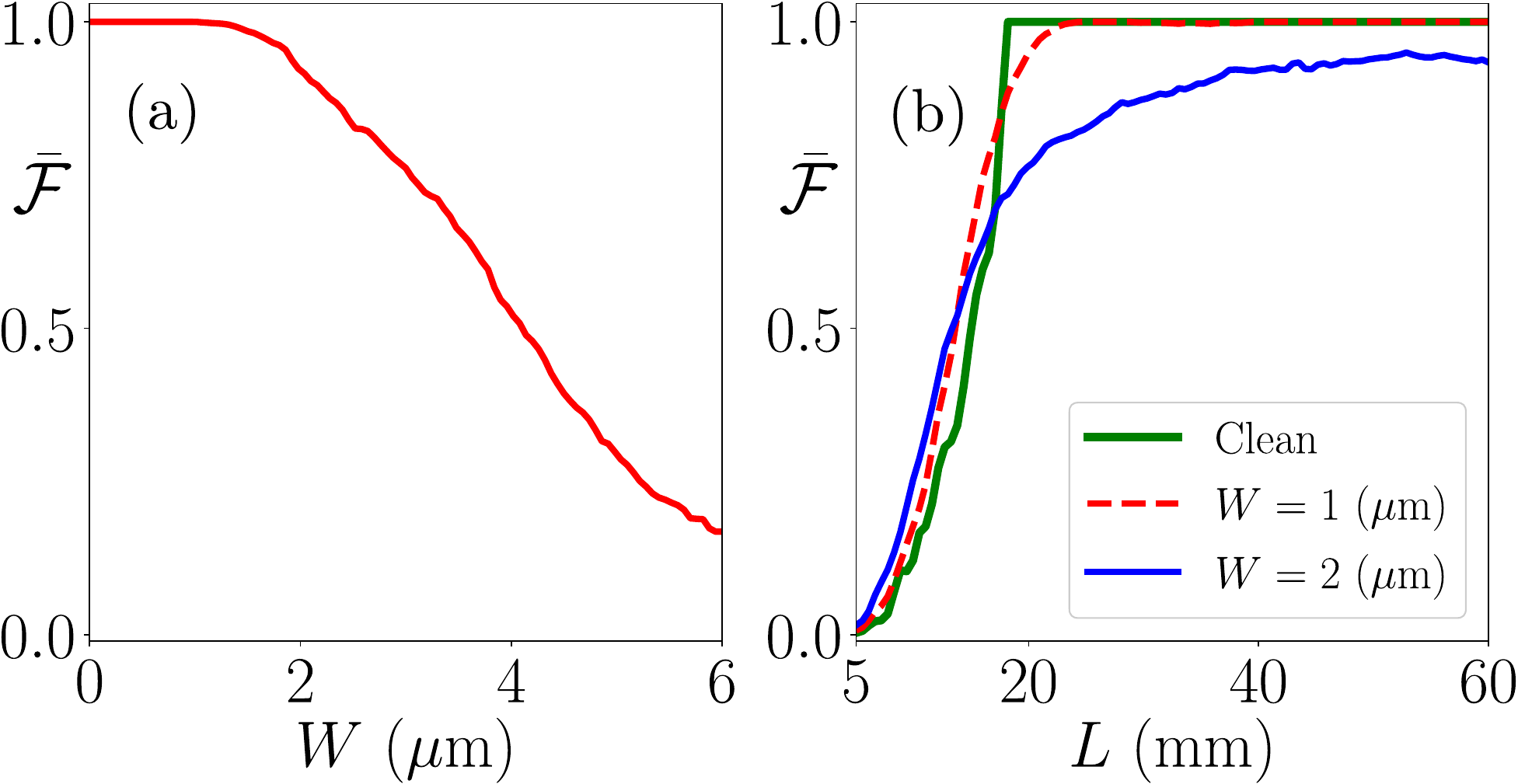}
	\caption{(a) Disorder-averaged output fidelity $\bar{\mathcal{F}}$ as a function of the disorder strength $W$ for a fixed total propagation length $L=40~\rm mm$. (b) Disorder-averaged output fidelity $\bar{\mathcal{F}}$ as a function of the total propagation length $L$ for the clean system (green), $W=1$~(red), and $W=2$~(blue). Here, $d_A=10~\mu \rm m$ and $d_B=7~\mu \rm m$. We take an average over $100$ random disorder configurations for each $W$.
	}
	\label{Fig:Disorder}
\end{figure}

We plot the disorder-averaged output fidelity $\bar{\mathcal{F}}$ in Fig.~\ref{Fig:Disorder}(a) as a function of the disorder strength $W$ for a constant total propagation distance $L$. We observe that the disorder-averaged fidelity $\bar{\mathcal{F}}$ remains close to $1$ for disorder strengths up to around $1.8~\mu \rm m$, which is around $20\mhyphen25\%$ of the clean inter-waveguide distances $d_{A,B}$. Further increasing the disorder strength leads to $\bar{\mathcal{F}}$ starting to decrease, which is expected in a strongly disordered system. We also show the disorder-averaged output fidelity $\bar{\mathcal{F}}$ as a function of the total propagation length $L$ for different disorder strengths in Fig.~\ref{Fig:Disorder}(b), with clean limit (green), $W=1~\mu \rm m$ (red), and $W=2~\mu \rm m$ (blue). For small to moderate disorder strengths, up to at least $W=1~\mu \rm m$, the disorder-averaged output fidelity $\bar{\mathcal{F}}$ remains almost the same as the clean system as a function of the total propagation length $L$, see the red and green curves in  Fig.~\ref{Fig:Disorder}(b). For stronger disorder, the output fidelity, however, does not reach 1 for the investigated total propagation lengths $L$, see the blue curve in Fig.~\ref{Fig:Disorder}(b). 
Taken together, the results in Fig.~\ref{Fig:Disorder} demonstrate that end-to-end pumping based on the Fibonacci-type waveguide array is robust against rather substantial random structural disorder.

\bibliographystyle{apsrev4-2mod}
\bibliography{bibfile.bib}

\end{document}